# Investigation of particle dynamics and classification mechanism in a spiral jet mill through computational fluid dynamics and discrete element methods

S. Bnà[1], R. Ponzini[1], M. Cestari[1], C. Cavazzoni[1], C. Cottini[2] and A. Benassi[2,3*]

1 - CINECA Supercomputing Centre, Via Magnanelli 6/3, 40033 Casalecchio di Reno (Italy)

2 - DP Manufacturing & Innovation dept., Chiesi Farmaceutici SpA, Parma (Italy)

3 - International School for Advanced Studies (SISSA), Trieste (Italy)

*Corresponding author address: Chiesi Farmaceutici S.p.A. Largo Belloli 11A– 43122 Parma (Italy)

email address: a.benassi@chiesi.com

Predicting the outcome of jet-milling based on the knowledge of process parameters and starting material properties is a task still far from being accomplished. Given the technical difficulties in measuring thermodynamics, flow properties and particle statistics directly in the mills, modelling and simulations constitute alternative tools to gain insight in the process physics and many papers have been recently published on the subject. An ideal predictive simulation tool should combine the correct description of non-isothermal, compressible, high Mach number fluid flow, the correct particle-fluid and particle-particle interactions and the correct fracture mechanics of particle upon collisions but it is not currently available. In this paper we present our coupled CFD-DEM simulation results; while comparing them with the recent modelling and experimental works we will review the current understating of the jet-mill physics and particle classification. Subsequently we analyze the missing elements and the bottlenecks currently limiting the simulation technique as well as the possible ways to circumvent them towards a quantitative, predictive simulation of jet-milling.

# 1 Introduction

Spiral jet mills are widely employed in many manufacturing sectors for comminution or deagglomeration of dry powders. The absence of any solvent or additive, the absence of mechanical moving parts and the simplicity of the device geometry made jet milling the preferred technique for particle size reduction in industries such as food and pharma where the control of contaminants and the ease of cleaning are pivotal. Compared to other milling techniques jet milling allows micron and sub-micron size reduction with relatively narrow particles size distributions, it is thus ideal for those applications and products extremely sensitive to small variation in size and physicochemical properties of the powder particles [1–3]. Despite its popularity the jet milling process is still far from being completely understood, at present no tool exists to predict quantitatively its outcome or to design and optimize it a-priori for a given powder and a given mill geometry. Thus most of the process design and scale-up activities are performed on trial and error basis with high costs, especially in the pharma industry where powders of newly synthesized drugs can reach a value of many tens of thousands euros per kilogram [4].

We believe in the possibility to build a computational tool based on the coupling of computational fluid dynamics (CFD) and discrete element methods (DEM) that could be helpful in both unravelling the basic physical mechanisms ruling the milling process and in designing custom processes and equipment for specific products as well as in assisting during process scale-up. In this long term perspective we hereby present a first set of CFD and DEM calculations on a simplified ideal model mill geometry and compare our findings with the other experimental and computational results available in literature critically evaluating some of the commonly adopted assumptions and approximations. The simulation campaign was not aimed at achieving a quantitative agreement with experimental data or to validate numerical models, rather to catch qualitatively the correct physics and to understand how to properly design the missing feature towards a predictive computational tool. We also examine in detail the scales and the orders of magnitude of several phenomena to highlight the possible bottlenecks both intrinsic in the CFD-DEM coupling method or dictated by computational power limits. Finally we comment about the next steps that should be accomplished in the direction of a full and self-consistent CFD-DEM description of the jet milling process. The paper is organized as follows: in the introduction section the basic physical principles of jet milling and the orders of magnitude involved are reviewed; in section two the CFD, DEM and coupling models and the assumptions therein are discussed in details, deeply technical aspects are treated in the Appendices; in section three the milling fluid dynamics is analysed as a function of geometry and process parameters based on the pure CFD simulations; sections four and five deal with the study of the classification mechanism through 1-way CFD-DEM coupling, a comparison is made with the predictions from the cut-size equation; section six describes the collision physics and statistics as a function of the powder hold-up; in section seven limitations and bottlenecks of the current model and future development perspectives are presented; finally conclusions are drawn in section eight.

## 1.1 Milling fluid mechanics

The sketch in Figure 1 (a) summarizes the typical jet mill working principle: a milling fluid, e.g. steam, air or nitrogen, is injected in the cylindrical milling chamber by several grinding nozzles whose upstream pressure $p_0$ is usually referred to as the *grinding pressure*. The number of nozzles is typically related to the size of the chamber, lab scale mills with a chamber diameter of 1 inch usually have no more than 4 nozzles while industrial scale mills 20-30 cm in diameter can reach up to 12 nozzles. For mineral and concrete/cement applications jet mills can reach up to 1-2 m in diameter. Nozzles are usually 1-2 mm in diameter so that, already at $p_0 \sim 2 - 3\ barg$ they reach the *critical conditions* (sometimes referred to as the *sonic choke*), i.e. the speed of the milling fluid at the nozzle throat locks to the speed of sound $v_{sound}$, for that fluid. Above the critical condition onset, the flow rate through the nozzles depends only on the upstream feeding conditions and is independent of the pressure and flow in the milling chamber. In this regime the milling fluid flow rate $\dot{m}_{max}$, its temperature $T_t$ and speed $v_t$ at the throat can be easily calculated using the isentropic flow theory [5]:

$$\dot{m}_{max} = A_t\, p_0 \sqrt{\frac{\gamma M}{RT_0}} \left(\frac{2}{\gamma + 1}\right)^{\frac{\gamma+1}{2(\gamma-1)}} \qquad (1)$$

$$T_t = T_0 \frac{2}{\gamma + 1} \qquad (2)$$

$$v_t = v_{sound} = \sqrt{\gamma R T_t / M} \qquad (3)$$

Where $p_0$ and $T_0$ are the upstream pressure and temperature respectively, $A_t$ is the nozzle throat section. $R$ is the ideal gas constant, $M$ the milling gas molar mass and $\gamma$ the specific heat ratio. Once in the milling chamber the milling fluid rotates forming a vortex towards the central outlet, this means the fluid elements velocity has both a tangential $v_t^f$ and a radial $v_r^f$ component with respect to the milling chamber centre, see Figure 1 (b). While it is not difficult to predict the behaviour of the milling fluid in the accelerating nozzles, its quantitative description into the milling chamber is far more complex and cannot be obtained without the aid of CFD. The sonic/supersonic nature of the milling fluid generates strong density and temperature gradients that can be correctly described only properly treating the mechanics and thermodynamics of compressible, non-isothermal, high speed flows.

1.2 Particle motion and particle-fluid interaction

The powder enters the milling chamber through a larger inlet dragged by other milling fluid, usually accelerated by a Venturi pump. To avoid blow back of milling fluid and powder the milling fluid pressure, usually referred to at the *feed pressure* $p_{feed}$, is typically kept $0.5 - 1.0\ bar$ larger than the grinding one. The powder particles are accelerated by drag and lift forces exerted by the fluid and their velocity vector can be decomposed into a tangential $v_t^p$ and a radial $v_r^p$ component, see Figure 1 (c). Small particles, whose inertia is negligible compared to the fluid drag and lift forces, will follow the fluid streamlines, large particles will give rise to more complex trajectories governed by their own inertia. The propensity of a particle to follow its inertia rather than the fluid streamlines is captured by the Stokes number [6]:

$$Stk = \frac{\rho_p p \delta^2 v_0}{18\ \mu\ D} \qquad (4)$$

Where $\rho_p$ is the particle density, $\delta$ its diameter, $\mu$ is the dynamic viscosity of the milling fluid, $v_0$ its velocity far enough from the particle so that its perturbation can be considered negligible and $D$ is a characteristic size of the milling chamber. For $Stk \ll 1$ inertia is negligible and the particle will follow strictly the fluid motion, for $Stk \sim 1$ or greater inertia will play a significant role in defining the particle trajectories. Assuming the particles are made of Lactose (a typical excipient for pharmaceutical inhalation products) $\rho_p = 1520\ Kg/m^3$, for $N_2$ the viscosity at ambient conditions is $\mu = 1.7 \cdot 10^{-5}\ Pa \cdot s$, the diameter of a typical pilot scale milling chamber is $D = 10\ cm$. As will be illustrated in the following sections $150\ m/s$ is a representative value for $v_0$. For large feed particles with $\delta = 100\ \mu m$ one gets $Stk = 74$ while for micronized particles with $\delta = 1\ \mu m$ one gets $Stk = 7.4 \cdot 10^{-3}$. Thus large feed particles move and collide mainly driven by their own inertia while finely micronized particles leave the milling chamber dragged by the milling fluid following perfectly its streamlines.

In a steady milling process a certain amount of powder is instantaneously present in the milling chamber kept in motion by the milling fluid (*hold-up mass*) [7]. Thus the milling fluid transfers energy and momentum to the powder particles accelerating them and simultaneously slowing down. As a consequence the fluid dynamics in the steady operation state is not only influenced by the grinding and feeding pressures but also by the powder feed rate $\dot{m}_{feed}$ which in turns determines the amount of hold-up $m_h$. Experiments show that the residence time $t_p$ of the particles in the milling chamber for a pilot scale mill fed at $2Kg/h$ ranges between $10 - 50\ s$ [7], the hold-up mass is related to the residence time by:

$$m_h = \dot{m}_{feed} t_p \qquad (5)$$

leading to an estimation of $5 - 30g$ of hold-up. A useful number allowing to characterize how complex is the particle-fluid interaction is the solid volume fraction $n$, i.e. the ratio of the volume occupied by the powder $V_p$ to the total volume of the milling chamber $V_p + V_f$:

$$n = \frac{V_p}{V_p + V_f} = \frac{m_h/\rho_p}{\pi D^2/4\ L} \qquad (6)$$

With $D$ and $L$ milling chamber diameter and thickness respectively. According to the literature [8] with $n < 10^{-6}$ the slowdown effect played by the powder on the fluid in negligible, with $n > 10^{-6}$ such effect becomes relevant. With $n > 10^{-3}$ indirect particle-particle interaction start to play a significant role, i.e. each particle feels the presence of the others due to their wake perturbation. Finally with $n > 10^{-2}$ direct particle-particle collisions are expected to be extremely frequent modifying drastically the particles trajectory. Assuming the same density and chamber size used to evaluate eq. (4) only the chamber height must be specified, a value of $L = 2\ cm$ gives a volume fraction between $0.02 - 0.1$. It is thus clear that the full phenomenology of particle-fluid and particle-particle interactions must be taken into account for a correct quantitative description of jet milling.

Despite the complexity of the particle-fluid interaction a simplified model to estimate the particle trajectories and the cut size diameter $\delta_{cut}$, i.e. the diameter below which the milled particles can escape the milling chamber through the outlet, can be built based on the balance between radial drag force of the fluid and centrifugal force of the particles [9–11]. The model is based on the decomposition of the particle motion in the radial and tangential directions in the x-y plane of Figure 1 (c), neglecting any lift term which would produce forces along the z-direction the particles will move along perfectly planar orbits with diameter:

$$r = \frac{4}{3} \frac{\delta}{C_D} \frac{\rho_p}{\rho_f} \left(\frac{v_t^p}{v_r^f}\right)^2 \qquad (7)$$

With $\rho_f$ fluid density and $C_D$ the drag coefficient for spherical particles, which depends on the Reynolds number $Re$ and thus on the particle diameter $\delta$ and on the relative velocity between the particle and the fluid. If, collision after collision, the particle diameter $\delta$ reduces the orbit of the fragments will shrink moving closer to the center of the milling chamber, i.e. closer to the classification rim and the outlet. Notice that, to obtain equation (7), the radial component of the general drag force expression, presented later in the manuscript in equation (11), has been approximated replacing the relative (or *slip*) velocity between particle and fluid $\overrightarrow{v^p} - \overrightarrow{v^f}$ with the radial component of the fluid velocity $v_r^f$. This corresponds to assume that the particle radial velocity is negligible compared to the fluid one.

Particles are *classified*, i.e. they leave the milling chamber when the radius of their orbit becomes equal or smaller than the classifier radius $d/2$. Thus substituting $d/2$ in equation (7) it is possible to obtain an expression for the cut size $\delta_{cut}$:

$$\delta_{cut} = \frac{3}{8} d \, C_D \frac{\rho_f}{\rho_p} \left(\frac{v_r^f}{v_t^f}\right)^2 \qquad (8)$$

Notice that in the last step the assumption $v_t^p = v_t^f$ has been made, this is true only if two conditions are satisfied:

1) the particles are very small so that $Stk \ll 1$, in this case the particles will follow the fluid stream lines with the same velocity;
2) The particle-particle collisions are not so frequent so that the fluid has enough time to accelerate the particles reducing the relative velocity to zero.

It has been already demonstrated that assumption 1) is reasonable for particles in the micron range as those expect to be classified. Assumption 2) holds only if the powder feed rate $\dot{m}_{feed}$ is small and the hold-up is reasonably small as well. Moreover the energy and momentum transferred by the fluid to the hold-up can slow down the fluid reducing $v_t^f$ [9–11]. If conditions 1) and 2) are fulfilled and the hold-up has a negligible effect both $v_t^f$ and $v_r^f$ can be estimated by CFD calculations and $\delta_{cut}$ predicted with eq. (8). The latter equation will be further discussed in the next sections comparing its predictions with the results of our simulations, to this aim it is useful to introduce the *spin ratio* as $v_t^f/v_r^f$.

When the hold-up mass is negligible or minimized a further increase in the grinding pressure $p_0$ does not lead anymore to a particle size reduction or sharpening of the product particle size distribution. In such conditions the *limit particle size distribution* is reached, its maximum value should coincide with $\delta_{cut}$. The shape and average value of the limit size distribution is a function of the powder material properties and mill geometry, it does not depend anymore on the process parameters and the size of the initial feed material.

A comparison between the predictions of equation (8) and the full 3D DEM simulations will be given in sections 4 and 5 giving us the opportunity to test the legitimacy of the assumptions and approximations above discussed.

### 1.3 Fracture mechanics and particle size reduction

Size reduction occurs by particle-particle and particle-wall collisions if, in between collisions, the particles collect enough kinetic energy being accelerated by the milling fluid. The size reduction can follow several different paths including simple breakage, chipping or fragmentation [12]. When a collision happens at too low energy the particles are bounced back elastically or they can be weakened without rupture (*fatigue*). The statistical description of particle rupturing upon collisions is based on the breakage and selection functions, both of which depends on the collision energy as well as on some material parameters such as the Young modulus $E$, describing particle elasticity, the Hardness $H$, describing particle resistance to plastic deformation and the Fracture Toughness $K_C$, describing the resistance to crack propagation [13–15]. Such material properties can be estimated through micro-indentation measurements: single particles can be cracked while recording the force versus penetration distance curves, from the shape of these curves $E$ and $H$ can be calculated. Measuring the length of the cracks departing from the indentation hole rim one can also estimate $K_C$ [14,16]. Atomic force microscopes can be also used to indent particles at the nano-scale [17–19], however, when probed at scales smaller than the characteristic disorder one, $E$ and $H$ are found to be scale dependent [20]. Moreover being the indentation hole smaller than single crystal grains $E$ and $H$ data present a large variability depending on which crystallographic orientation is exposed locally by the surface. It is only with indentation holes of several microns that the material response is averaged over the smaller scale disorder and becomes less variable and scale independent. Another way of measuring the Hardness of particles is by direct compression of a powder bed through the Heckel model [21–23], coming from the collective response of the entire powder bed, the $H$ value is already averaged over all the possible crystallographic orientations of the powder particles and over many possible particle-particle contact geometries.

1.4 Material properties and their alteration

Material properties play a major role in determining the outcome of a jet milling process for at least two main different reasons. On one hand the particle resistance to breakage and the crack propagation depend on the material crystallographic properties and on its purity [21,24–26]. On the other hand the fracturing process itself can modify the solid state properties at the fragments surface, one of the most striking problems with pharmaceutical powders is the surface amorphization [27–29]. Being the amorphous state thermodynamically meta-stable at ambient conditions the particle surface will gradually reconvert to its crystalline state by thermal activation on a time scale that ranges from hours to weeks. This behaviour can modify the drug product performances in time and its stability, during re-crystallization the milled particles can form solid bridges and get fused together with a drastic increase of the particle size distribution in time. Alteration of the material properties are usually avoided or at least minimized by optimizing the milling process or by applying a post-milling conditioning to control and accelerate the re-crystallization. For those pharmaceutical powders showing a thermodynamically stable amorphous state, milling induced amorphization could be beneficial as it modifies the drug dissolution profile usually enhancing its bioavailability [30].

From equation (8) it is also evident that the molecular weight of the milling gas itself can be used to modify the milling efficiency, this possibility is discussed by many authors [9,31].

The shape of particles can also influence the milling process, as discussed above it enters directly into equation (8) through the drag coefficient. But irregular shapes and surface roughness can also influence the way particle break upon collisions [10,32].

1.5 Modelling and experiments

While the great deal of complexity described in the previous sections still prevents the milling process from being completely understood and mastered, empirical or semi-empirical approaches to the scale-up of the process across different milling plants have been put in place sometimes with successful results. The most popular is probably the one based on the conservation of the specific milling energy [31,33], another approach is based on the dimensionless number derived by Mueller et al. [7] representing the grinding condition to maintain constant while scaling the process. More recently approaches based on CFD simulations have been adopted to estimate $\delta_{cut}$ based on more sophisticated models than the one leading to equation (8) [9,11]. Such models require the knowledge of $v_t^f$ and $v_r^f$ at the classifier rim and their behaviour as a function of the milling chamber geometry and process parameters, such quantities are provided by the CFD calculations. The inclusion of the hold-up effects comes through additions empirical parameters to be measured experimentally. Recently Rodnianski et al. [34] pushed this approach to the limit calculating all the necessary fluid and collision properties through CFD-DEM simulations, parametrizing their dependence on geometry and process parameter as well as on the hold-up. This massive computational effort allowed the construction of a light and quick statistical model, a sort of population balance model, able to predict in certain cases with remarkable agreement, the full experimental particle size distribution. Other authors adopted CFD-DEM approaches to study the particle dynamics inside the milling chamber [35–39], the statistics of collisions and their distribution [40,41]. The implementation of collision models [42] and of smart numerical solutions to speed up the simulations [43] have been documented, however a self-consistent CFD-DEM description of jet milling at realistic time scales and powder feed rates is still far from being achieved.

Very interesting, in the perspective of validating computational models for jet milling, is the possibility to image and measure fluid and particle velocities through particle image velocimetry [44–46]. Such experiments allow the visualization of the supersonic plum out of grinding nozzles and to map the particle and fluid velocity field in the milling chamber.

## 2 Model description
### 2.1 The mill geometry

The present computational study has been performed on a model (simplified) milling chamber geometry capturing all the main features of real bottom discharge pilot scale plants. The principal sizes and dimensions are reported in Table 1 and illustrated in Figure 1 (d).

| Geometry element | Symbol | Values | Units |
|---|---|---|---|
| milling chamber diameter | $D$ | 100 | mm |
| milling chamber height | $L$ | 16 | mm |
| outlet classifier diameter | $d$ | 20,35,50 | mm |
| outlet classifier penetration in the milling chamber | $\ell$ | 9.5,12.75 | mm |
| powder inlet diameter | - | 6.5 | mm |
| powder inlet insertion angle | $\beta$ | 30 | deg |
| powder inlet insertion distance | $a$ | 20 | mm |
| number of nozzles | - | 6 | - |
| nozzle length | - | 10 | mm |
| nozzle diameter | - | 1 | mm |
| nozzle angle | $\alpha$ | 26,40,50 | deg |

Table 1: Numerical values for sizes and geometrical features of the simulated mill geometry.

During the study several geometrical features has been varied to assess their impact on the cut-size and more generally on the milling process physics. More specifically the effect of changing the nozzle angle $\alpha$ with respect to the milling chamber walls, the penetration of the classifier rim into the chamber volume $\ell$ and the diameter of the classifier itself $d$ have been evaluated.

## 2.2 The CFD model

The milling gas steady state is described through an Eulerian approach and has been calculated using the open source CFD software OpenFOAM® [47,48]. The following assumptions have been made in the choice of the solver and boundary conditions:

- Compressible ideal gas: given the high Mach numbers expected, the compressible nature of the milling fluid must be taken into account;

- Turbulent flow: given the high Reynolds numbers expected, the turbulent nature of the milling fluid flow cannot be neglected;

- Isentropic flow at the nozzles: given the small length of the grinding nozzles and the high speed of the nitrogen passing through them the assumption of non-dissipative (reversible), adiabatic (no heat exchange) flow, i.e. isentropic flow, is certainly acceptable. The nitrogen flow is subsonic in the upstream reservoir feeding the mill and in most of the milling chamber as well, however passing through the small diameter cylindrical grinding nozzles it accelerates to the sound speed reaching and locking to the critical condition. Just outside the nozzle it experiences a sudden expansion becoming supersonic in the next 2-3 mm and thus cooling down.

- Non-isothermal flow: the sudden expansion of the fluid in sonic/supersonic conditions causes large temperature drops, even down to 150 °K. In such conditions the energy conservation equation must be coupled with the Navier-Stokes one.

According to these assumptions the rhoSimpleFOAM solver has been adopted; it belongs to the pressure-based segregated solvers family and it implements the SIMPLE algorithm originally proposed by Patankar et al. [49,50]. While the original version of the solver has been demonstrated to work properly for most of the subsonic flow, these segregated methods have been later extended to compressible flow problems including transonic or supersonic flows. The efficiency of the segregated methods in the case of high speed flows is usually worse with respect to the so-called density based solvers [51]. However, for applications like our one where the flow is transonic only in a small region of the integration domain localized at the nozzle outlets, the TRANSONIC variant of the SIMPLE solver has been demonstrated to converge properly. Technical details about the solver used and its numerical convergence can be found in Appendix A.

Turbulence is not simulated explicitly but included in a Reynolds average fashion (RANS approximation) through an effective viscosity accounting for the energy dissipation by the turbulent eddies. For the effective viscosity a simple $k - \varepsilon$ model has been adopted in analogy with what has been done in most of the CFD literature concerning jet mill simulation [50] ($k$ represents the turbulent kinetic energy per unit mass of the fluid and $\varepsilon$ the rate of dissipation of such energy respectively). Such a model is numerically robust, cheap from the computational point of view and reasonably accurate, however the eddy-viscosity models are insensitive to the streamline curvature, swirl and rotation and this could result in a non-optimal description of cyclones, swirls and vortexes [52,53]. Corrections are available as well as more accurate ways of including turbulence effects, such as the more complicated Reynolds Stress Model (RSM), but at the price of a higher calibration and computational costs. Technical data about the turbulence model parameters are detailed in Appendix A.

The gas feeding of the mill enters the simulations through the boundary conditions, the latter can be fixed according to the typical values directly measured in the plant. Usually in a real pilot plant the only accessible physical quantities upstream the milling chamber are the grinding and feeding pressures, the corresponding gas volumetric flow rates and the gas temperature. Downstream, a few centimetres away from the classifier pipe, the gas is already at atmospheric pressure and close to ambient temperature. No probe can be easily placed inside the milling chamber to monitor pressure, temperature or flow rates during the process both because its presence might alter the normal flow path of the milling gas and because the presence of the powder could lead to its abrasion and damaging. Table 2 shows typical values measured on a pilot plant like the one idealized in our simulations:

| Feeding pressure $p_{feed}$ [barg] | Grinding pressure $p_0$ [barg] | Upstream temperature $T_0$ [˚C] | Downstream temperature $T_{out}$ [˚C] | Feeding flow rate $\dot{m}_{feed}$ [$Nm^3/h$] | Grinding flow rate $\dot{m}_{max}$ [$Nm^3/h$] |
|---|---|---|---|---|---|
| 6±0.05 | 5±0.05 | 23 | 31.5 | 8.5±1 | 20.5±1 |
| 7±0.1 | 6±0.1 | 23 | 34.2 | 9.5±1 | 26.5±1 |
| 8±0.1 | 7±0.1 | 24 | 38.7 | 10.5±1 | 30.0±1 |
| 9±0.1 | 8±0.1 | 23 | 39.3 | 11.5±1 | 36.2±1 |
| 10±0.1 | 9±0.1 | 22 | 42.2 | 12.5±1 | 41.1±1 |

Table 2: Typical operational conditions for a pilot scale jet mill

Notice that the usual prescription of keeping the feeding pressure 0.5 to 1 bar higher than the grinding one to avoid blow back has been adopted [31]. The upstream gas temperature is measured by a built-in sensor before the flow splits in the feed and grinding circuits, the downstream temperature has been measured with a thermocouple placed in contact with the outer surface of the metallic classifier pipe just outside the milling chamber.

At the six nozzle inlet surfaces, green in Figure 2 (a), a constant mass flow rate boundary condition has been imposed. The mass flow rate is fixed according to the values in Table 2 (last column) divided by a factor 6 to account for the flow splitting through the six nozzles considered to be equivalent. The gas velocity and pressure at the nozzle entrance are calculated during the simulation, their value can be used to calculate the total energy per unit mass $u$ at the nozzles and compare it with the value measured upstream the mill as a consistency check. Figure 2 (b) shows such comparison for different grinding pressures, the energy per unit mass has been calculated as follow:

$$u = \frac{e}{\rho} + p\,V + \frac{v^2}{2} = h + \frac{v^2}{2} = C_p\,T + \frac{v^2}{2} \qquad (9)$$

where $e$ is the internal energy per unit volume, $p$, $V$ and $v$ the generic pressure, volume and velocity of the gas, $C_p$ the constant pressure specific heat and $T$ the gas temperature. It is assumed no energy loss due to viscous forces and no localized pressure drops between the upstream measurement point and the nozzle entrance. While the former assumption is reasonable as only few centimetres separate the two points of interest, the latter might be too strong as the gas flow experiences some abrupt change in the piping diameters and some sudden turn before entering the six nozzles. This could be the cause of the non-perfect match of the two energy values in Figure 2 (b), however a discrepancy of 6% is perfectly acceptable for the purpose of the present calculation, especially considering that the real geometry differs from the simulated one. The same strategy has been adopted for the powder feed inlet, blue in Figure 2 (a). For the milling chamber outlet, the red patch in Figure 2 (a) at the end of the classifier pipe, a constant pressure boundary condition has been adopted verifying that, by the mass conservation principle, the total outgoing gas mass flow rate equates the total incoming gas mass flow rate from the inlets. An example of such check is shown in Figure 2 (c) where the experimental mass flow rate is plotted together with the imposed flow rate at inlets and the calculated flow rate at the outlet for different grinding pressures. A perfect matching between the incoming and outgoing mass flow rates is a further check of the good convergence of the numerical solutions. In most of the calculations the outlet pressure has been set to ambient pressure (1 atm). On all the other inner walls, according to the boundary layer theory, a no slip boundary condition has been applied. A zero gradient value is prescribed for the pressure in all the patches except the outlet one to which an outflow condition is prescribed for the velocity field.

The turbulent fluid behaviour close to the milling chamber walls must be accurately described in the calculations, the thickness of the boundary layer $q$ can be estimated using the flat plate theory knowing the fluid velocity far from the walls (free stream velocity) and setting a characteristic length for the fluid-wall interface [5]. Using a free stream velocity of 250 m/s and the milling chamber perimeter as the characteristic length, the thickness of the boundary layer results to be $q = 300 - 400 \mu m$. Such thickness should be finely sampled during the volume discretization with a mesh size $\Delta x_{CFD} \ll q$ leading to huge meshes and thus unaffordable computational costs. Alternatively *wall functions* can be introduced in the calculations for the turbulence fields $k$ and $\epsilon$ so that the size of the mesh elements closer to the walls can remain large, comparable in size with the boundary layer itself [50]. More details about the calculation of the boundary layer and the wall function approach can be found in Appendix A. A fixed value of $k$ and $\epsilon$ is prescribed on the inlet surfaces, to determine such value the analytical formula valid for the turbulent flow in pipes has been employed [54]. Lastly, a zero gradient for $k$ and $\epsilon$ is prescribed on the outlet section.

Having no possibility to measure the gas temperature inside the milling chamber a fixed value of 20ºC has been applied on the inner chamber walls based on the evidence that the thin metallic walls outside the chamber remain at ambient temperature even after many ours of milling operation. For the fluid incoming from the powder feed inlet the upstream temperatures reported in Table 2 have been adopted. At the grinding nozzles the temperature has been set using equation (2) with $T_0$ being the upstream temperatures reported in Table 2 , i.e. the critical condition has been assumed. Such assumption can be readily verified on the pilot plant by measuring the grinding mass flow rate $\dot{m}_{max}$ while changing the pressure in the milling chamber (via changes in the feeding pressure $p_{feed}$), keeping fixed the grinding pressure upstream the nozzles $p_0$. For a nozzle at the critical condition equation (1) predicts that the gas mass flow rate depends only on $p_0$ and $T_0$ and not on the downstream conditions. Indeed looking at Figure 2 (d) the measured grinding mass flow rate remains constant upon changing $p_{feed}$ and scales linearly with $p_0$ as predicted by (1), panel (e). Thus, down to $p_0 = 2$ bar, the grinding nozzles operate always in the critical condition no matter how $p_{feed}$ is chosen.

The geometry of the real plan just out of the classifier rim differs from the simulated one that has been deliberately kept simpler, as a consequence the downstream temperature $T_{out}$ reported in Table 2 cannot be directly applied as a boundary condition at the red patch of Figure 2 (a). An adiabatic boundary condition has been applied instead, the temperature that freely sets at the outlet has been compared with $T_{out}$ in Figure 2 (f). This is the only case where the simulation results do not compare with the measured data showing an outlet temperature which is basically independent of the pressure while it should grow linearly with it. This discrepancy is certainly dictated mostly by the difference between the real and the simulated geometries and for the sake of a qualitative understanding of the milling process it is perfectly acceptable. Surely for future simulations aiming at a quantitative agreement with the measured data a better treatment of the temperature at the boundaries is necessary as well as an accurate way of measuring it directly at the mill outlet.

The meshing strategy adopted to discretize the integration volume is discussed in Appendix A together with a mesh sensitivity analysis showing which is the minimum grid size necessary to achieve a steady state whose properties become independent on the mesh size itself.

2.3 One-way CFD-DEM coupling

The coupling with the DEM description of the particle motion is performed through the software LIGGGHTS® [55]. It is an unresolved one-way coupling, i.e. the fluid exerts a force on the particles but not vice-versa, so the solid mass cannot slow down the fluid which remains in its steady state calculated through the CFD simulations. We deal essentially with pure 3D DEM simulations where the presence of the fluid is accounted by drag and lift forces acting on the particles and varying locally according to its density, temperature and velocity field. Being the coupling unresolved the fluid volume elements must be bigger in size compared to the particle diameter and inside each fluid volume element temperature, density and velocity are constant. Particles passing simultaneously through the same fluid volume element can experience different drag and lift forces as these forces depend on the relative velocity $\vec{v^p} - \vec{v^f}$ and on the relative angular velocity $\vec{\omega^p} - \vec{\omega^f}$.

As a reference particle density we have chosen the lactose one $\rho_p = 1525\ Kg/m^3$. For both the particle-particle and particle-wall interactions the Hertz-Mindlin model has been selected, a contact history for the tangential force component has also been included [56]. A constant direction torque model has been used for the rolling friction and no cohesion has been included [57]. More details about the particle-particle interactions and the model coefficients can be found in Appendix B.

The first aim of our one-way CFD-DEM coupling is to verify the validity of the cut size model leading to equation (8). In this respect we start by including only the drag force in the particle equations of motion:

$$\vec{F}_{drag} = \frac{\pi}{12} C_D \rho_p \delta^2 \left(\vec{v^p} - \vec{v^f}\right)^2 \frac{\vec{v^p} - \vec{v^f}}{\left|\vec{v^p} - \vec{v^f}\right|} \qquad (10)$$

with the drag coefficient given by the empirical correlation of Schiller-Naumann [58]:

$$C_D = Max\left[0.44, \frac{24}{Re_p}\left(1 + 0.015\ Re_p^{0.687}\right)\right] \qquad (11)$$

$Re_p = \rho_f \left(\vec{v^p} - \vec{v^f}\right) \delta/\mu$ is the particle Reynolds number. This form of the drag force on a single spherical particle is correct only for dilute solid phase, i.e. volume fractions $n < 10^{-3}$, above such threshold the spherical solid particles do not travel anymore in an unperturbed fluid, they feel each other wake and this usually increases the drag force they experience [8]. Di Felice-like corrections can be included to account for this effect, both for mono- and poly-disperse particles, but they require a continuous evaluation of the solid volume fraction which is computationally expensive, in this first study we neglected them. Corrections exist also to account effectively for the non-sphericity of the particles [8]. It must be stressed that, whatever degree of accuracy one puts in the evaluation of the drag force, equation (10) and its variants are still not adequate for small particles travelling close to the milling chamber walls. Here the particles meet the boundary layer, i.e. the region of the fluid where the velocity decreases reaching zero exactly at the wall. We account for the existence of this region implicitly through the wall functions thus, in our simulations, particles close to the walls feel an average velocity value which departs more from the correct one the smaller and the closer the particles are to the walls. One might argue that for our kind of applications wall functions are not an appropriate choice, however even resolving explicitly the boundary layer with a mesh refinement at walls does not work. According to the estimation made in the previous section the boundary layer is roughly $q = 300 - 400 \mu m$, this means mesh elements must be at least $\Delta x_{CFD} = 20 - 50 \mu m$ to resolve it accurately and ensure a good numerical convergence. With such small mesh size an unresolved coupling is possible only with particles whose diameter is smaller than $\delta = 5 - 10 \mu m$, larger particles could not be simulated at all. Provided that a mixed resolved-unresolved coupling in a multi-phase flow is far from being formulated and implemented in a high performance computing software, we believe that our choice represent the best compromise to model the jet mill physics. Still this is a crucial point to be solved as only a correct description of the forces felt by particles close to walls can lead to the correct impact velocity during collisions and thus to the correct description of particle breakage.

As the particle density is much larger than the fluid one virtual mass and pressure gradient forces have been deliberately neglected. Given the large fluid velocities drag and lift forces are expected to play a major role thus gravitational and buoyancy forces have been neglected as well. With these further simplifications and disregarding for the moment any lift and torque contribution the equations of motions become:

$$\frac{d\overrightarrow{v^p}}{dt} = -\frac{3}{4}\frac{\rho_f}{\delta\,\rho_p}\,C_D\left|\overrightarrow{v^p}-\overrightarrow{v^f}\right|\left(\overrightarrow{v^p}-\overrightarrow{v^f}\right) + \left(\overrightarrow{v^p}-\overrightarrow{v^f}\right)\cdot\widehat{r^p}\,\frac{\overrightarrow{r^p}}{|\overrightarrow{r^p}|} + \frac{F_{col.}}{\pi/6\,\rho_p\delta^3} \quad (12)$$

$$\frac{d\overrightarrow{\omega^p}}{dt} = \frac{T_{col.}}{\pi/60\,\rho_p\delta^5} \quad (13)$$

$\overrightarrow{r^p}$ is the particle position vector and $\widehat{r^p}$ its versor, fixing the origin for such vector field in the center of the milling chamber helps in projecting the slip velocity components along the radial and tangential directions. $F_{col.}$ and $T_{col.}$ are the total force and torque acting on each particle that is colliding with one or more other particles. Notice that the two equations are coupled through the collision terms, i.e. only through the collisions energy can be transferred between translational and rotational degrees of freedom. Neglecting the collision term while searching a stationary solution for equation (12), i.e. imposing $d\overrightarrow{v^p}/dt = 0$, leads precisely to equation (8).

The DEM technique itself has two main limitations both evident when applied to jet milling problems:

- Working with small particles requires the time integration step $\Delta t$ to be small as well, usually $\Delta t \sim 20\%$ of both Rayleigh and Hertz time to ensure a good resolution of each collision event [8]. To fulfil these conditions when simulating $\delta = 1\ \mu m$ particles one has to set $\Delta t = 10^{-9} s$ with evident limitations in simulating seconds or even minutes of milling process dynamics.

- Working with high speed poly-disperse particles also forces the time integration step to be very small. Large particles reach a larger terminal velocity and in a timestep $\Delta t$ they could travel distances longer than the diameter of smaller particles, these results in missing particle-particle collisions or in numerical instabilities if a large and a small particle interpenetrate each other. Moreover large and fast particles can cross the milling chamber walls if $\Delta t$ is not small enough to resolve particle-wall collisions. To simultaneously simulate a poly-disperse solid phase with particle diameters from 1 to $50\ \mu m$ a timestep $\Delta t = 10^{-9} s$ is again necessary.

These limitations can be overcome at the price of renouncing to the correct description of collisions and angular velocity of particles. Neglecting the collision term and the rotational degrees of freedom equation (12) becomes:

$$\frac{d\overrightarrow{v^p}}{dt} = -\frac{3}{4}\frac{\rho_f}{\delta\ \rho_p}\ C_D \left|\overrightarrow{v^p} - \overrightarrow{v^f}\right|\left(\overrightarrow{v^p} - \overrightarrow{v^f}\right) + \left(\overrightarrow{v^p} - \overrightarrow{v^f}\right)\cdot\widehat{\overrightarrow{r^p}}\ \frac{\overrightarrow{r^p}}{|\overrightarrow{r^p}|} \qquad (14)$$

It is now evident that this equation of motion depends on the product $\delta\ \rho_p$ only: the particle trajectories will remain the same upon changing arbitrarily $\delta$ and $\rho_p$ provided that their product remains constant. The trajectories of lactose particles with diameter $\delta$ can be obtained by simulating larger particles of fake dimeter $\delta_{fake}$ with a smaller density $\rho_{fake} = \rho_p\ \delta/\delta_{fake}$. With such a trick injection of poly-disperse particles with diameters differing by two or more orders of magnitude can be simulated; larger particles allow to keep $\Delta t$ larger, even up to $5\cdot 10^{-8} s$ with no numerical stability problems. Panels (a) and (b) of Figure 3 show the superimposition of true (black) and fake (orange) particle trajectories for subsequent time frames in the case of $\delta = 1\ \mu m$ particles, ascending the classifier rim, and $\delta = 20\ \mu m$, colliding with the vertical chamber wall. The particle position and behaviour are indeed the same in both true and fake density/diameter cases. Notice how in the fake case the particle injection takes longer: being $\delta_{fake} \gg \delta$, it is not possible to inject the particles with the same rate in the same injection volume, this problem can be easily circumvent enlarging the injection volume region. Another interesting difference to notice for the $\delta = 20\ \mu m$ case is that the real particles, while orbiting, tend to occupy the bottom corner of the milling chamber while the fake ones form a thicker ring occupying part of the vertical wall as well. Again this is due to the large value of $\delta_{fake}$ which prevent the particles from getting closer "condensating" at the edge.

This kind of DEM simulations are useful to verify the cut-size value in different CFD steady states obtained varying the process parameters or the milling chamber geometry. However having particles with a diameter $\delta_{fake}$, usually bigger than the real one, the collision energy and frequency are not correctly predicted as well as the powder volume fractions. Still, poly-disperse particle injections can be used to understand if and how the milling fluid vortex separates particles of different size, where the particle-particle and particle-wall collisions are more abundant and if these regions are different for different particle size. Panel (c) and (d) of Figure 3 emphasizes one of the main limitations of the technique, it compares the calculated kinetic and rotational energy of mono-disperse particles with $\delta = 1, 20$ and $100\ \mu m$ in the two cases of true and fake diameter method. As long as they are injected $1\ \mu m$ particles try to reach the classifier spreading over a large volume, injecting only 0.1 million particles their density is thus very small and both particle-particle and particle-wall collisions are negligible (see also snapshots of panel (a)). In such conditions the particle kinetic energy is 2 orders on magnitude larger than the rotational one, energy exchange between the translational and rotational degrees of freedom is also negligible leading to a perfect match of the kinetic energies for the true and fake density/diameter methods. After a longer simulation time part of the particles not able to reach the classifier remain trapped in the milling chamber rotating in its periphery, panel (e) and (f) show how the particle-wall collisions are distributed in such steady state. The fake diameter method reproduces quite well the true region where the collisions occur and their "surface density" what is slightly overestimated is the collision velocity. The true particles have a very small diameter and tend to move closer to the horizontal chamber walls frequently colliding with them, as a result their average speed is small. On the contrary fake diameter particles are larger, the volume close to the walls get crowded and more particles must move away from it, where the fluid velocity is higher and its drag can accelerate the particle stronger. This picture is confirmed by the collision probability, i.e. the number of collisions occurring in a time instant divided by the total number of particles present in the milling chamber in that instant. These data are presented in Table 3 for particle-particle and particle-walls collisions with $\delta = 1, 20$ and $100 \, \mu m$ for both the true and fake methods. In all cases the number of collisions has been calculated during the steady state reached by the powder long time after the injection, all the data are generated with 0.1 million particles as the total number of particle is known to affect significantly the collision rate and type [40]. For the $1 \, \mu m$ case the number of particle-wall collisions is strongly decreased compared to the true diameter case, the large encumbrance of fake diameter particles prevent most of the particles to reach the chamber walls and collide. Upon injection, $20 \, \mu m$ particles reach immediately the vertical walls of the milling chamber "condensating" at the bottom edge (see snapshots of panel (b)). Here the powder volume fraction grows significantly and a large number of particle-particle and particle-wall collisions occurs. As a result the true density/diameter method kinetic and rotational energies differ by less than one order of magnitude, now the energy transfer between translational and rotational degrees of freedom become significant. Being the fake density/diameter collisions badly described the fake rotational kinetic energy is wrong by more than one order of magnitude leading to a discrepancy between true and fake kinetic energy with is also almost one order of magnitude. Table 3 shows how, having $\delta_{fake} \gg \delta$, a larger number of collisions takes place in the fake case than in the true one, with consequent decrease in both kinetic and rotational energy, however the fake kinetic energy is larger than the true one. This apparent contradiction originates from the fact that a larger number of collisions promotes the fake particle spreading over larger volumes, the rightmost figure of panel (b) shows how the ring of fake orange particles is wider than the true black ones. If particles move in a different region of the chamber where the milling fluid is rotating faster this could lead to an increase in their average velocity despite the number of collisions they originate is also larger. Finally in the $100 \, \mu m$ case kinetic and rotational energy are comparable in magnitude, here however $\delta_{fake} \approx \delta$ and the imprecision in the description of collisions vanishes. Although the fake rotational energy is still underestimated by one order of magnitude, the two kinetic energies are again in good agreement.

|        |      | Particle-wall collision probability | Particle-particle collision probability |
|--------|------|-------------------------------------|-----------------------------------------|
| 1 µm   | fake | 0.17                                | 0.10                                    |
|        | true | 0.73                                | 0.00                                    |
| 20 µm  | fake | 0.26                                | 2.43                                    |
|        | true | 0.17                                | 0.33                                    |
| 100 µm | fake | 0.42                                | 1.38                                    |
|        | true | 0.33                                | 1.07                                    |

Table 3: Interparticle and particle-wall collisions for true and fake methods applied to different particle diameter injections.

Results of poly-disperse particle injections on different CFD steady states will be shown in the following sections, in all of them $\delta_{fake} = 200 \mu m$ has been used and $\rho_{fake}$ chosen to represent particles of 1,2,5,10,20 and 50 $\mu m$ simultaneously. Usually 0.1 million particles are injected with 4.5 $g/s$ mass flow rate. A timestep $\Delta t = 5 \cdot 10^{-8} s$ has been used and no particle breakage has been allowed.

If a more comprehensive description of particle motion is desired lift forces should be added to equation (12) and Stokes torque should be added to equation (13). The Magnus lift force is obtained as:

$$\vec{F}_{mag} = \frac{\pi}{8} \rho_f \delta^3 \left[ \left( \overrightarrow{v^p} - \overrightarrow{v^f} \right) \times \left( \overrightarrow{\omega^p} - \overrightarrow{\omega^f} \right) \right] \qquad (15)$$

with $\overrightarrow{\omega^p} - \overrightarrow{\omega^f}$ relative rotational velocity between particle and fluid (the rotational velocity of a fluid is defined as twice the fluid vorticity field). This formulation of Magnus lift assumes implicitly a Rubinow-Keller correlation for the lift coefficient [59]. The Saffman lift force is expected to be negligible as the particles are very small compared to any geometrical feature of the system, velocity gradients are thus expected to take place on a scale much larger than the particle diameter. The only exception is represented again by large particles moving close to the milling chamber walls and feeling the presence of the turbulent boundary layer.

The Stokes torque (or stokesian rotational drag) for low Reynolds numbers can be written as:

$$\vec{T}_{stokes} = \pi \mu \delta^3 \left( \overrightarrow{\omega^p} - \overrightarrow{\omega^f} \right) \qquad (16)$$

More complex correlations exist for high particle Reynolds numbers and non-spherical particles. The effect of the inclusion of lift and torque contributions into the equation of motion will be discussed in the last part of the present work in terms of particle energy, trajectories and collision properties.

For computational efficiency the density, temperature and velocity field calculated with CFD on a non-structured, non-uniform mesh with average step $\Delta x_{CFD}$ are remapped on a coarser structured cubic mesh with step $\Delta x_{DEM}$ before being passed to the DEM code for particle trajectory integration. The choice of $\Delta x_{DEM}$ is known to influence the calculated particle dynamics if $\Delta x_{DEM} < 1.4\ \delta$ [34], in most of the CFD-DEM coupling codes it is recommended to keep $\Delta x_{DEM} \sim 2 - 3\ \delta$ [60]. A mesh sensitivity study is performed in Appendix B.

## 3 Milling gas dynamics

In this section the milling gas behavior at fixed working conditions is described and discussed starting from the flow at the grinding nozzles. A 2D map of the gas velocity magnitude inside and in the vicinity of a grinding nozzle is presented in figure 4 (a) for a grinding pressure $p_0 = 10\ bar$, relevant thermodynamic quantities along the white AB segment are plotted in panel (b). In the interior of the cylindrical nozzle the fluid is slightly subsonic, it reaches the sonic condition almost at the nozzle exit. If the nozzle would continue indefinitely keeping the same cylindrical diameter the critical condition would lock the gas velocity to the sound speed, however the presence of the larger milling chamber causes a sudden expansion of the sonic gas resulting in its acceleration above the speed of sound. During such supersonic expansion density and pressure drop down significantly in few millimeters only. In keeping expanding the gas starts decelerating eventually becoming subsonic again. The temperature profile is also interesting, in the initial subsonic and the subsequent supersonic acceleration the gas cools down according to the isentropic flow theory [5] reaching -125 °C. When decelerating, both during the initial supersonic part and the final subsonic one, the gas heats up again, this is the reason why the gas temperature at the milling chamber outlet is higher than upstream, this heating effect increases with increasing upstream pressure, see Figure 2 (f). Notice how the supersonic plum bends with respect to the nozzle entrance direction due to the swirl flow present in the milling chamber. This phenomenon has been reported by other authors performing CFD simulations [10,11,37,61] and observed experimentally by particle image velocimetry [45].

The radial and tangential components of the milling fluid velocity for the working condition $p_0 = 8\ bar$ are plotted in Figure 5 on three different planes: the horizontal one (x-y plane) crossing the nozzles (a) and (d); the horizontal one at the classifier rim (b) and (e); the vertical one oriented 90º from the powder feed inlet (c) and (f). These planes are also highlighted in panel (a) of Figure 6 in orange, blue and green respectively. The radial component of the velocity has a complex behavior changing sign and varying by one order of magnitude within few millimeters, to highlight all the features of this field in a single picture a saturation of the color scale has been necessary. Looking at panel (a) of Figure 5 a black halo is evident all around the outer boundaries of the milling chamber, here the velocity points outwards favoring the particle-wall collisions and thus the size reduction. However if a particle is attracted too close to a supersonic plume (where the velocity points inwards, orange color) its gets bounced back towards the milling chamber. This finding is in line with the commonly agreed picture of the supersonic plumes behaving like impenetrable walls promoting comminution of those particles unable to make their way around them [61]. Moving towards the center of the milling chamber the radial component of the velocity remains positive, pushing the particles outwards, but it experiences a sudden drop by one order of magnitude, this is also visible in panel (b) of figure 6. Only in the central part of the chamber and close to the classifier the color switches to orange, i.e. the radial velocity points inwards (the lines in panels (a), (b) and (c) are $v_r^f = 0$ iso-lines highlighting the change in sign). Only in this region the drag force points in the right direction to compensate the centrifugal one and circular orbits are possible, i.e. only here the model leading to equations (7) and (8) for the cut size is valid. Panel (b) of Figure 5 shows how the situation does not changes significantly moving from the nozzle to the classifier rim plane, the black halo is attenuated and almost vanishing. That the latter is localized only around the nozzle plane is visible in panel (c) showing the $v_r^f$ behavior along the milling chamber height. The same panel also show how the orange regions, where circular orbits are possible, are localized on the classifier rim and on the horizontal chamber walls, see also panel (c) of figure 6 for a 2D profile. If small particles and fragments are produced by collisions with the vertical outer walls, the only way they have to cross the milling chamber, reach the classifier and leave, is through to the horizontal walls. As qualitatively illustrated also by other authors [38,61] the classification mechanism rely on complex 3D particle trajectories spending most of their time in the close vicinity of walls. Nevertheless the prediction of the cut size $\delta_{cut}$ through eq. (8), i.e. through the simple 2D orbit model, works nicely as it makes no assumption on the particle motion and forces along the direction perpendicular to the orbit. Clearly the asymmetry of the isolines and the whole $v_r$ field with respect to the angular coordinate is due to the presence of the powder feed inlet, a similar condition has been obtained in other CFD modeling works whenever the powder feed inlet has been explicitly simulated [36].

The tangential component of the velocity $v_t^f$ is more uniform along the milling chamber circumference, along its thickness and it varies smoothly and monotonically moving along the radial direction. Being always positive it is supposed to push the particles to move counterclockwise everywhere. 2D maps are plotted in panels (d), (e) and (f) of figure 5 while 2D profiles are illustrated in figure 6 (d) and (e). Inhomogeneities in the $v_t^f$ distribution are expected to appear as a result of the hold-up if the powder density is not uniform inside the milling chamber during milling at high feed rates. In the plotted profiles $v_t^f$ is almost constant while approaching the classifier edge, however, depending at which height with respect to the classifier they are taken, they can also exhibit an increase as the one captured by the particle image velocimetry experiments [46].

Figure 6 displays in panels (f) and (g) the inverse of the spin ratio $v_t^f/v_r^f$ along the three lines depicted in panel (a), according to the 2D orbit model this quantity is connected to the cut size $\delta_{cut}$ through eq. (8). As already discussed only where $v_r^f$, and thus the spin ratio, has a negative value the drag force can compensate for the centrifugal force and circular orbits can exists. The larger the spin ratio the smaller $\delta_{cut}$, i.e. only smaller particles can be classified. The spin ratio will be studied as a function of the process parameters and the milling chamber geometry in the next sessions.

We conclude the session analysing the behaviour of the other physical quantities describing the fluid behaviour in the milling chamber. These are plotted in Figure 7 along the same lines of the velocity components. Examining the profiles along the EF direction it is immediately clear that pressure, density and temperature are almost constant along the milling chamber thickness. Moving from the classifier rim to the outer walls both fluid pressure and density increase almost linearly, only the temperature shows a non-linear growth and its value close to the outer walls is extremely position dependent due to the presence of the six supersonic plumes.

## 4 Influence of the process parameters on the cut size

As a further step the grinding pressure $p_0$ has been modified from 6 to 10 bar keeping the feeding pressure $p_{feed}$ always one bar more to prevent blow-back phenomena. The analysis of the fluid velocity and its relevant thermodynamics quantities are summarized in Figure 8. The qualitative behaviour already illustrated in the previous section does not change: the radial component of the velocity, panel (a), remains almost unaffected by the pressure change close to the outer walls, it grows almost linearly with pressure only in the vicinity of the classifier rim. The tangential component of the velocity, panel (b), grows linearly with $p_0$ everywhere, this should lead to stronger particle acceleration, more energetic collisions, eventually enhancing the grinding efficiency. As both the velocity components grow linearly the spin ratio, panel (c), remains unaffected by variations in feeding and grinding pressure, this finding is confirmed by the CFD analysis performed by Rodnianski et al. [11]. Also the temperature grows linearly with the grinding pressure while density and internal pressure gradients are made more steep but, approaching the classifier rim, they reach the same value: the cut size is solely determined by the spin ratio.

To better capture and quantify the particle classification physics as a function of the milling fluid properties equation (8) has been solved using density, temperature and velocity values taken along the classifier circumference, $0.5\ mm$ above the rim, i.e. the black dashed line in the inset of Figure 9 (a). Equation (8) must be solved numerically as the drag coefficient $C_D$ is on its part a function of $\delta_{cut}$ through the particle Reynolds number as visible in equation (11). The slip velocity appearing in $Re_p$ has been replaced by the radial fluid velocity $v_r^f$ consistently with the approximations introduced in deriving equations (7) and (8). The necessity to employ the fluid velocity and density fields in the calculation of eq. (11) appears immediately by the $Re_p$ definition, the temperature field enters in the estimation of the viscosity $\mu$ according to the Sutherland model. The numerical solution of (8) is plotted in Figure 9 (a): the cut size is non-uniform along the classifier circumference due to the asymmetry introduced by the powder feed inlet, i.e. the largest particles able to escape the milling chamber will do it crossing the classifier rim in a specific position of limited length, crossing elsewhere is not permitted; very small particles with $\delta \ll \delta_{cut}$ can cross the classifier rim everywhere. The largest value for $\delta_{cut}$ is roughly 0.5-0.6 µm independently of $p_0$, this value can slightly change if calculated at a different height with respect to the classifier rim but, even estimating it few fractions of mm away, does not exceed 1 µm. Only particles with diameter 1 µm or smaller can be collected out of the mill regardless the $p_0$ value. This finding is against any experimental evidence as typically, increasing $p_0$ (at constant powder feed rate), leads to a reduction of $\delta_{cut}$ until the limit distribution is reached. We can conclude that particle classification, and ultimately the size of the product powder, is mainly driven by the milling fluid slowdown caused by the hold-up, the only phenomenon that is not accounted for by the simple theory leading to eq. (8). Higher pressures increase $v_t^f$ and thus the particle collision energy and the milling efficiency, this means a smaller residence time of the particles in the milling chamber and, at fixed powder feed rate, a smaller hold-up amount. With smaller hold-up the speed $v_t^f$ is further increased while $v_r^f$ remains unaltered, this unbalances the spin ratio lowering $\delta_{cut}$ as $p_0$ grows. The necessity to model implicitly or to simulate explicitly the fluid slowdown with 2-way coupling CFD-DEM, in order to catch the correct dependence of $\delta_{cut}$ from $p_0$, has been demonstrated by other authors [9,34]. Other works presenting CFD-DEM simulations with 1-way coupling only, i.e. not including by definition the possibility for the powder to slow down the fluid, reported the impossibility of simulating the classification of particles larger than 1 µm [38].

The same situation is found upon modification of the outlet pressure $p_{out}$, the results for $\delta_{cut}$ are shown in panel (b) of Figure 9. Any modification of the pressure downstream the milling chamber, maybe due to the opening/closing of powder collection bowls or safety valves should not be able to affect the classification mechanism.

That $\delta_{cut} \sim 1\ \mu m$ can be also demonstrated explicitly by injecting polydisperse particles on the CFD steady state with the 1-way coupling previously described, an example is presented to conclude the section for the $p_0 = 7\ bar$ case of figure 9. Particles with $\delta = 1 \div 50\ \mu m$ are injected from the powder feed inlet at an average rate of 4.5 $g/s$ with equal probability distribution in mass. The injection stops once 0.1 million particles are introduced in the simulation box, the whole injection takes roughly 10 $ms$ as visible from panel (a) of figure 10. Panel (b) of the same figure shows the particle positions few instants after the injection started, size segregation is clearly visible with larger particles taking larger orbits closer to the milling chamber vertical walls. Panel (a) also shows the outgoing mass flow rate through the classifier which is solely composed by $\delta = 1\ \mu m$ particles and ceases slightly after the injection stops. From the delay between the two mass flow rate curve it is possible to estimate the residence time for 1 $\mu m$ particles to be roughly 5 $ms$. Once the injection is completed only a limited amount of 1 $\mu m$ particles can escape the milling chamber before a steady state is reached, panel (c) of figure 10 shows the top and side view of the particle distribution inside the milling chamber in such steady state. 1 $\mu m$ particles are orbiting allover the roof of the chamber, where the radial component of the velocity is inward oriented, 2 $\mu m$ particles have the same behaviour but they remain segregated on the bottom wall of the chamber. Larger particles, whose orbit radii would be bigger than the chamber one, are confined to the edges where the largest volume fraction is reached and where most of the particle-particle and particle-wall collisions take place. All the particles avoid the central region of the milling chamber where the supersonic plumes are located, this condition is clearly sustainable only as long as the hold-up mass is small. More details about the particle dynamics and the collision dynamics as a function of $\delta$ and hold-up mass will be given in the next sections.

## 5 Influence of the mill geometry on the cut size

The analysis described in the previous section demonstrated how robust is the cut size value against variation of the input and output pressures. It is also possible to test its robustness against geometry variations, we will concentrate in particular on the effect of the nozzle entrance angle $\alpha$, on the penetration depth $\ell$ and diameter $d$ of the classifier. Increasing $\alpha$ the supersonic plumes protrude more towards the center of the milling chamber where they are still bent by the vortex flow, an example of how this behavior effects radial components of the velocity is given in panels (a) to (c) of Figure 11, the color scale is the same of Figure 5 allowing for a direct comparison of the velocity maps. The formation of a grinding halo (or comminution zone) around the outer chamber walls upon increasing $\alpha$ is clearly visible: on the nozzle plane, panel (a), the large orange ring represents a region where large particles can be trapped into circular orbits until collisions reduce their size. Collisions are promoted just above and below the nozzle plane, panel (b), where the halo is black, i.e. $v_r^f$ points outwards pushing particles against the chamber walls. The onset and the sharpening of the grinding halo with increasing $\alpha$ is better appreciable plotting $v_r^f$ along the chamber radius like in panel (d). The existence of such grinding halo has been previously reported experimentally by particle image velocimetry [45,46], it has been found to enlarge with decreasing nozzle number while its behaviour as a function of $\alpha$ is in qualitative agreement with our simulations: increasing the entrance angle the fluid velocity in the halo reduces (the tangential component mainly), while the halo width remains substantially unaltered.

Comparing panels (c) of both Figure 5 and 11 it is evident how $v_r^f$ is strongly affected also in the central part of the milling chamber, rising $\alpha$ the regions where the velocity points inwards increase significantly: fine enough particles can reach the classifier rim more easily and not necessarily sneaking close to the walls. This does not necessarily mean that larger particles can escape easily and that the cut-size increases, to infer information about $\delta_{cut}$ the same calculation leading to the plots of Figure 9 must be performed on the classifier circumference. The spin ratio itself, panel (f) of figure 11, depends significantly on $\alpha$ only far from the classifier. The $\delta_{cut}$ value along the classifier rim is shown in Figure 14 (a) and indeed no significant variations are found as a function of $\alpha$. We are thus lead to the conclusion that increasing the nozzle entrance angle might increase the comminution efficiency, by both increasing the grinding halo volume and the tangential component of the fluid velocity (panel (e) of Figure 11), but should not affect significantly the classification. This finding is in agreement with recent measurements by Luczak et al. [46] showing no difference in the grinding performance at two different $\alpha$ values and different flow rates. Notice, however, that both our findings and the measurements by Luczak et al. are in contrast with the older work by Katz and Kalman [62] providing evidence of a size reduction of the classified particles with increasing $\alpha$. As for the results of the previous section the cause of the discrepancy must be attributed to the effect of the hold-up: an increase in $\alpha$ leads to higher $v_t^f$ values, i.e. to an enhanced grinding efficiency and hold-up reduction, thus to a further increase in $v_t^f$ and consequent shift of the spin ratio and $\delta_{cut}$ to lower values. In favour of this thesis are the CFD-DEM simulations by Han et al. performed with a 1-way coupling, i.e. not allowing for fluid slowdown, where the nozzle orientation was found to produce no significant change on the simulated outcoming particle size distribution.

Another important geometry element whose role is often discussed in literature is the classifier pipe. We modified both its penetration height $\ell$ and its diameter $d$. Increasing $\ell$ from 9.5 to 12.75 mm has only a little effect on $\delta_{cut}$ which is found to increase by a small fraction of micron, see panel (b) of figure 14. 2D maps of the radial component of the fluid velocity on the plane perpendicular to the milling chamber are shown in panels (a) and (b) of Figure 13 for the two $\ell$ values simulated, no significant difference is observed in the two cases. These results are in agreement with the findings of Kozawa et al. [38] using a 1-way coupling CFD-DEM approach and with the CFD analysis performed by Rondniansky et al. [11]. Experimental evidences confirm our finding of a slight increase of $\delta_{cut}$ with increasing $\ell$ [38,63], however the magnitude of such increase is larger: few microns rather than fractions of micron, the reason is again the missing slowdown exerted by the powder on the fluid.

Finally, keeping $\ell = 12.75\ mm$, we modified the classifier diameter $d$: the impact of such modification is shown in Figure 12 in terms of radial and tangential components of the fluid velocity as well as on the spin ratio. Reducing the size of the classification pipe increases the fluid velocity in the radial direction, especially close to the classifier rim. Simultaneously a slight reduction of the tangential velocity is observed. 2D maps of the radial component of the fluid velocity on a plane perpendicular to the milling chamber can be compared looking at panels (b) to (d) of Figure 13. The major differences are located at the classifier rim and close to the external classifier walls, no differences exist in outer region of the chamber as well as close to the supersonic plumes. The spin ratio, Figure 12 (c), is now significantly different for the three simulated classifiers, however when multiplied by $d$ according to eq. (8), the differences get attenuated and $\delta_{cut}$ results indeed different but still very limited in range between 0.5 and 1.5 μm, see Figure 14 (b). A value of $\delta_{cut} \sim 2\mu m$ can be found if the calculation is performed slightly above the classifier rim. Again even changing the classifier diameter no significant change in the cut size, i.e. in the distribution of the product material, is predicted by the model neglecting the fluid slowdown caused by the hold-up.

Like for the pressure dependence of $\delta_{cut}$ analysed in the previous section also here the cut size calculations can be confirmed with DEM particle injections. An example is given in Figure 15 for the case $p_0 = 8\ bar$, $p_{feed} = 9\ bar, p_{out} = 1\ atm$ with $\alpha = 50°, d = 50\ mm$ and $\ell = 12.75\ mm$ which is expected to allow the classification of the largest particles. Indeed injecting particles in the same conditions described before for Figure 10 both 1 and 2 $\mu m$ particles are now able to leave the chamber while the injection is still ongoing. Upon stopping the injection 2 $\mu m$ particles remain trapped in the chamber while 1 $\mu m$ particles continue to flow out emptying completely the chamber. Panel (b) shows the behavior of the different particle populations, interestingly now 2 $\mu m$ particles rotate close to the roof of the chamber while 5 $\mu m$ particles populate its bottom, particles with larger diameter behave as before.

6 Other considerations on particle dynamics and collision statistics

In the previous section lift and torque terms described by equations (15) and (16) have been deliberately neglected to comply with the assumptions made in the derivation of the cut-size equation (8). To understand how important are these fluid-particle interaction terms for the realistic description of the particle behavior we switched on them one by one in DEM simulations performed over the reference steady CFD state with conditions $p_0 = 8\ \text{bar}, p_{feed} = 9\ \text{bar}, p_{out} = 1\ \text{atm}, \alpha = 50°, d = 50\ \text{mm}, \ell = 12.75\ \text{mm}$. Different injections with real diameter particles have been performed evaluating the impact of lift and torque on the classification mechanism of 1 $\mu m$ fine particles as well as on the collision statistics of 20 $\mu m$ particles, the results are shown in Figure 16. In panel (a) it is possible to see that no significant modification in the classification mechanism occurs, the outgoing mass flow rate and the residence time of $\delta = 1\ \mu m$ particles remains the same. The same applies for the probability distributions of the particle-particle and particle-wall collision velocity reported in panels (b) and (c) for large particles orbiting in the periphery of the milling chamber. Evidently both lift and torque contributions, scaling like $\delta^3$, play a minor role compared to the drag term which scales like $\delta^2$, this is at least true for particles having diameter up to many tens of microns.

Panels (b) and (c) of Figure 16 also show how particle-wall collisions occur on average at larger relative velocities than the particle-particle ones, in fact walls are at rest while particles are moving all in the same direction with similar velocities. One could be thus lead to the conclusion that particle-wall collisions are the main responsible of size reduction, however this holds only in our RANS approximation. Local and time dependent fluctuations of the velocity field caused by the turbulent eddies, and averaged out in the RANS approach, could modify the particle trajectories, e.g. making them more chaotic, thus increasing the probability of high energy particle-particle collisions. How to incorporate the effect of turbulence on the particle trajectories, as well as why it is meaningless to do it in a one-way coupling, is discussed in section 7. Finally it has to be noted that, for large powder feed rates, large amounts of hold-up will crowd the milling chamber and some of the particles will not be able to circumvent the supersonic plumes out of the grinding nozzles. In a collision with a particle the supersonic plumes will behave like a wall promoting grinding, such situation has already been observed experimentally in the early works by Kürten and Rumpf using triboluminescence [64]. This particle breakage mechanism, directly induced by the plumes, might alter the collision statistics and energetics only at large feed rates or in low milling efficiency situations where the powder hold-up is significant.

A last point which is interesting to address is the effect of increasing the number of particles in the milling chamber. To this aim the previously described injections of 100k mono-disperse $20~\mu m$ particles have been compared with 500k and 1M particle injections. With small particle numbers most of the collisions occur with the milling chamber walls, only 20% of them are between particles, as the particle number increases the situation is reversed: most of the collisions are between particles and only 20% is between particles and walls. This behavior is illustrated in Figure 17 (a), an analogous tendency has been reported by Dogbe et al. [40]. In agreement with the latter work we also noted that the particle-particle collision velocity decreases with increasing particle number, see panel (c) of Figure 17. This is due to the shortening of the mean free path and mean free time between particles collisions leaving less time for the fluid to re-accelerate them. Figure 17 (b) shows instead how the distribution functions for the particle-wall collision velocity remain unaltered by the increase of particles number. Thus:

- Increasing the particles number the particle-wall collisions are inhibited and the milling efficiency as well. Increasing the powder feed rate can lower the milling efficiency not only by slowing down the milling fluid but also by crowding the milling chamber walls thus reducing the occurrence of very energetic particle-wall collisions.

- The particle-particle collisions do not slow down the particles, they occur in fact at very small relative velocity and between particles moving all in the same direction. Thus when eventually particles meet the milling chamber walls the collision velocity remains high and unaffected by the number of injected particles.

- The particle-particle collisions just slightly deflect the particle trajectories randomly reducing the particle-wall collision frequency.

These conclusions are supported by the scatter plots of Figure 17 (d) and (e) showing the normal and tangential components of the collision velocities in both particle-particle and particle-walls collisions. As already noted by Teng et al. [41] the tangential component is predominant revealing how particles collide sidewise while keep moving in the same direction. Increasing the particles number the clouds of points get larger revealing a more chaotic behavior still with the tangential component prevailing on the normal one. Finally the picture above illustrated is confirmed by the spatial distribution of the collisions in panel (f) for 100k, 500k and 1M particle injections. The black points, locating particle-particle collisions, grow in density in the vicinity of the milling chamber wall as the number of injected particles grows, this region collects all the particles temporarily "distracted" from their run against the wall.

## 7 Critical considerations on the model and future improvements

The results of our CFD study are in line with those presented in the currently available literature although the technical details for such simulations, e.g. mesh size and kind, solver used and numerical integration schemes, treatment of the boundary layer, are almost never detailed by most of the authors. The simulated milling fluid behavior is physically sound and partially validated by direct measurements on a pilot plant whose geometry is similar to the simplified one here employed. Large is the room for improvements:

- A density based solver could be used to better describe the steady shocks at the boundary of the supersonic plumes;

- As mentioned in Section 2.2 a simple eddy-viscosity model does not describe accurately the velocity distribution of swirling flows. Many possible steps toward a better description of the convection term could be taken although very demanding, sometimes prohibitive, from a computational point of view. To the best of our knowledge all the CFD studies of jet mills presented so far in the literature stick to the $k - \varepsilon$ model while on other subjects, e.g. cyclone design [52,53], medical devices for inhalation and aerosolization design [65–67], the choice of a turbulence model capable of accurately describe vortexes and swirling flows is quite debated.

This said, we still believe such points are of secondary importance compared to the modifications the swirling flow velocity profiles could experience due to the presence of the powder hold-up. Unfortunately, in the few works available in literature that features a 4-way coupling, the effect of the hold-up mass on the fluid is never illustrated explicitly and no velocity profiles with and without powder are available. Further improvements in the CFD model will be reconsidered only after the implementation of a 4-way coupling.

The DEM model has been so far intentionally kept as simple as possible. Aiming at a quantitative agreement with experimental results the model should be calibrated, the impact of the different parameters on the overall simulation outcome evaluated, the contact and non-contact interaction potentials accurately chosen. However, to be able to simulate realistic powder feed rates of few Kg/h, a coarse-graining approach is necessary. To quantify the number of particles and collisions produced during the milling process consider the following example: assume the coarse powder feed is composed by 100 $\mu m$ particles and that the particles are classified only when their size reaches 1 $\mu m$, this means that every incoming particle must be reduced to $10^6$ fragments. Assuming that the result of every collision is the splitting of the mother particles into two identical fragments this means that every feed particle give rise to $2 \cdot 10^6$ collisions. If the powder feed rate is taken to be $1 \, Kg/h$ one has $3.5 \cdot 10^5$ particles entering the milling chamber every second (assuming the true density of the powder to be the lactose one), in the steady milling condition the same mass of 1 $\mu m$ particles must leave the mill i.e. $3.5 \cdot 10^{11}$ fine particles cross the classifier every second. Such huge number of particles is not manageable by any state-of-the-art high performance computing DEM code. The calibration of the DEM model against experimental measurements must be postponed after the evaluation and implementation of a proper coarse-graining strategy. Such strategy, aiming at representing a large group of particles by means of a single simulated one, must:

- Replicate the correct collision frequency, the correct proportion between particle-particle and particle-wall collisions, the correct collision velocity probability distributions. A wrong collision statistics would in fact lead to a completely fictitious breakage statistics;
- Incorporate a particle breakage model coherent with the assumptions made in designing the coarse-graining;
- Replicate the correct particle trajectories, especially for small particles close to the classifier rim, otherwise particle classification will be totally unrealistic.

The particle-fluid coupling also demands some attention, two are at present the main missing ingredients:

- The empirical correlation (10) does not include any indirect particle-particle interaction, necessary when realistic powder volume fractions are achieved. As already mentioned in Section 2.3, Di Felice-like corrections are available and will be included in the forthcoming 4-way coupling implementation;
- During their permanence inside the milling chamber, particles spend most of their time close to walls. As discussed in Section 2.3 and Appendix B, a non-resolved coupling cannot account for the fluid velocity drop experienced by particles as they enter the fluid boundary layer. Empirical correlations exist to account for the drag and lift forces variations when a particle is close enough to a wall/plate [68], they are however valid for single particles, to the knowledge of the authors extensions of theses correlations for the case of large powder volume fractions have never been published;
- The non-explicit treatment of turbulence in the CFD simulations prevent the possibility for the particle trajectories to be influenced by the multi-scale, time-dependent eddies. Which is the impact of such approximation in jet milling applications is hard to estimate a priori, both the particle speed and the Reynolds number are in fact high. A well known solution to correct the particle trajectories, implicitly accounting for the missing turbulent fluctuations in the fluid velocity field and thus on the drug force, is to add a Langevin-like stochastic term to the particle equations [69]. Again it makes sense to implement such advanced corrections only when a 4-way coupling with a proper coarse-graining approach will be able to represent the correct RANS fluid velocity filed in presence of a realistic amount of powder in the mill.

One encouraging evidence is that lift and torque contributions are in first place negligible, at least for particles up to 20-50 µm, this certainly simplifies the coarse-graining procedure allowing for the use of a fake density-diameter pair as described in Section 2.3.

## 8 Conclusions

With our one-way coupling CFD-DEM simulations we have shown and discussed:

- The behavior of the velocity, temperature and density profiles of the milling gas as a function of the milling pressure and of the main geometric features of the milling chamber.

- How complex is the real path of fine particles through the classifier outlet. While orbiting these particles reach the classifier rim moving close to the chamber walls, these are in fact the only regions where the radial velocity of the milling fluid is directed inwards towards the center of the milling chamber. The fact that fine particles prefer to move along the walls could explain the very weak or absent dependence of the cut-size diameter from the chamber height $L$ reported in the experiments [62].

- How the cut-size equation (8) works nicely in predicting the correct maximum size of the classified particles despite the simplistic assumptions made in its derivation. We have shown how the fluid radial velocity points inward, opposing to the particle centrifugal force, only close to the milling chamber walls and in a narrow halo just above the classifier rim. In most of the milling chamber volume, on the contrary, the fluid radial velocity points outward and no circular orbits for the particle are possible.

- How robust is the classification mechanism against significant variations in the milling chamber geometry and process parameters. This indicates the fluid slowdown, caused by the powder hold-up, as the major responsible for the variations in particle classification when the milling pressure is reduced or the powder feed rate is increased. This also calls for the implementation of a 4-way coupling and a coarse-graining approach to be able to capture the hold-up effect within the simulations.

- How the drag force is the most important of the particle-fluid interactions terms for particles below 50 $\mu m$, neglecting all the other terms a simplified equation allow for the correct description of the particle trajectories using a fake, larger, particle diameter. This numerical trick allows to simultaneously simulate particles with 2-3 orders of magnitude difference in their diameter and could be exploited in the design of a smart coarse-graining approach.

- How the collision energy and frequency changes with increasing hold-up, and how relevant are the particle-particle and particle-wall collisions. This information is also fundamental in the design of a coarse-graining approach able to correctly catch the collision statistics.

The analysis presented in this work constitutes a first step towards the implementation of a computational tools for the design and scale-up of jet milling processes, the current bottlenecks and the further steps to circumvent them have also been discussed.

# 9 Acknowledgments
AB is grateful to Riccardo Rossi from Red Fluid Dynamics for helpful discussions about CFD turbulence modelling and particle-fluid interaction. AB also acknowledge the technical support from DCS Computing on the DEM software and on the one-way CFD-DEM coupling.

# Appendix A: details about the CFD simulations

Table 4 summarizes the parameters used to calculate the milling fluid stationary state with the rhoSimpleFOAM solver and the $k - \varepsilon$ turbulent model:

| Feature | Value |
| --- | --- |
| Transonic flag | yes |
| Consistent | no |
| pMinFactor | 0.05 |
| pMaxFactor | 20 |
| residualControl | $10^{-6}$ |
| Solver | GAMG |
| Smoother | Gauss-Seidel |
| Solver tolerance | 1e-07 |
| RelaxationFactors | Fields:<br>  p  0.5<br>  rho  0.01<br>Equations:<br>  p = 0.9<br>  U, k, $\varepsilon$ and T = 0.55 |
| Turbulence model | $k - \varepsilon$<br>$C_1 = 1.44$<br>$C_2 = 1.92$<br>$C_3 = 0$<br>$\sigma_\mu = 0.09$<br>$\sigma_k = 1$<br>$\sigma_\varepsilon = 1.3$ |

Table 4: Details for rhoSimpleFOAM solver settings and $k - \varepsilon$ turbulence model

The constants used in the simulations to represent the Nitrogen milling gas properties are presented in Table 5 below.

| Property | Value | Units |
| --- | --- | --- |
| Constant pressure specific heat* $C_p$ | 1.04 | kJ/(kg °K) |
| Specific heat ratio* $\gamma = C_p/C_v$ | 1.4 | / |
| Molecular weight | 28.013 | Da |
| Density* $\rho_f$ | 1.251 | Kg/m³ |
| Viscosity model $\mu(T)$ | Sutherland law | |
| $A_s$ | $1.40673 \cdot 10^{-6}$ | Pa s / °$K^{1/2}$ |
| $T_s$ | 111 | °K |
| $T_0$ | 300.55 | °K |
| $\mu_0$ | $1.781 \cdot 10^{-5}$ | Pa s |

Table 5: Constants for the description of the milling fluid (* at STP conditions, i.e. 0°C and 1 atm)

The Sutherland law for the temperature dependence is implemented in its 3 parameters form in the LIGGGHTS® code:

$$\mu(T) = \mu_0 \left(\frac{T}{T_0}\right)^{\frac{3}{2}} \left(\frac{T_0 + T_s}{T + T_s}\right) \quad (17)$$

and in the 2 parameters form in OpenFOAM®:

$$\mu(T) = \frac{A_s T^{\frac{3}{2}}}{T + T_s} \quad (18).$$

The meshing strategy adopted to discretize the integration volume requires the three following steps:

- Generation of background cubic mesh of the overall bounding box using the standard OpenFOAM® tool blockMesh;
- Definition of the features lines of the reference geometry using the standard OpenFOAM® tool surfaceFeatureExtract;
- Definition of all the internal cells, projection the internal cells faces into the nearest cad surface and refinement on specific cad surfaces when needed by using the using the standard OpenFOAM® tool snappyHexMesh.

In this way high quality meshes can be obtained. Moreover, since the background cubic mesh is the driving parameter to define the grid size keeping constant the meshing topology in terms of level refinement ratios, it is very easy to generate a set of meshes with different size and perform a mesh sensitivity analysis to define the minimum size requested to solve the reference CFD model problem at hand. To this aim the same steady state calculation ($p_0 = 7\ bar$, $p_{feed} = 8\ bar$, $\alpha = 26°, d = 35\ mm$ and $\ell = 9.5\ mm$) has been repeated on five different meshes containing 12, 15, 24, 28 and 34 million elements approximately, the pressure drops between the outlet and the grinding /feeding inlets has been compared. Panels (a) and (b) of Figure A.1 show two regions of the finer and coarser meshes employed, panel (c) of the same figure shows the pressure drops, revealing a low mesh sensitivity and confirming the high quality of the 24 million cells mesh which has been adopted for all the simulations presented in the paper.

The turbulent boundary layer developing close to the milling chamber walls has been treated through wall functions. With this approach the smallest mesh elements close to the walls must have a size comparable to the boundary layer thickness $q$, a rough estimation of $q$ can be obtained using the flat plate model if a value for the fluid free stream velocity is available. Once $q$ is estimated a size for the wall mesh cells can be introduced in such a way that the dimensionless boundary layer thickness $y^+$ has a value between 30 and 300. The equations can now be solved and the correct value for the free stream velocity obtained. With this new value $q$ and $y^+$ can be adjusted iteratively. As a starting guess for the velocity we used the sound speed in Nitrogen gas and we corrected it iteratively, the calculation of $q$ requires also to estimate the length scale for the fluid-wall contact, given the swirling shape of the fluid velocity field, we used the perimeter of the milling chamber.

## Appendix B: details about the DEM simulations

The parameters for the particle-particle (pp) and particle-wall (pw) interactions are summarized in Table 6.

| Property | pp value | pw value | Units |
|---|---|---|---|
| Young modulus $E$ | 5 | 100 | MPa |
| Poisson ratio $\nu$ | 0.45 | 0.45 | / |
| Restitution coefficient | 0.2 | 0.2 | / |
| Sliding friction coefficient | 0.5 | 0.5 | / |
| Rolling friction coefficient | 0.3 | 0.1 | / |

Table 6: DEM contact model parameters

The chosen Young moduli are orders of magnitude smaller than the real lactose and steel ones, their value has been lowered to keep the Rayleigh time small thus allowing for large integration timesteps. Being less stiff both particles and walls will experience larger deformations but the same particle-particle and particle-wall forces will be generated during the simulations. This is a standard trick applied in most of the DEM simulations and it is known not to affect the results as long as the simulated powder particles are not dense and do not undergo a strong compression, like e.g. in tabletting simulations or ball milling simulations [70]. In jet milling simulation the dilute nature of the powder should allow the use of such approximation with no further worries. The choice of the other contact model parameters has been driven by our experience in modeling lactose powders, however they result from the calibration of DEM simulations for static and dynamic applications far from the range of densities and energies involved in jet milling. The correct setting of such parameters will be possible only when CFD-DEM simulations will be comparable with experimental data. We believe the proposed parametrization is still acceptable for the sake of our preliminary qualitative study, in most of the cases the selected values are comparable with the available literature on CFD-DEM modelling of jet milling.

The steady state condition of the milling fluid, described by the velocity, temperature and density fields calculated through CFD, are passed to the DEM code to compute the drag and lift forces on the particles, in such step the CFD mesh is coarsened from a small $\Delta x_{CFD}$ to a larger $\Delta x_{DEM}$. A lower bound for $\Delta x_{DEM}$ is known to be $1.4 \div 2\,\delta$ in order for the assumptions of the unresolved coupling to be fulfilled [34,60], LIGGGHTS® requires $\Delta x_{DEM} \geq 3\,\delta$. Multiple injection simulations have been run with $\Delta x_{DEM} = 1, 2.5$ and $5\,mm$ which means from $5\,\delta_{fake}$ to $25\,\delta_{fake}$ to search for an upper bound of $\Delta x_{DEM}$. The results concerning the classification of small particles are shown in Figure B.1 (a): with 1 and 2 $mm$ meshes only 1 $\mu m$ particles are partially classified, a small variation in the residence time is present; the 5 $mm$ case shows a complete classification of 1 $\mu m$ particles. This significant change in the classification behavior is due to the strong under-sampling of the milling gas radial velocity profile, panel (b) of Figure B.1 shows the almost continuous profile from the CFD calculations and the three discretized profiles coming from our choices of $\Delta x_{DEM}$, 2D maps of the same quantity on the classifier rim plane are shown in panel (c). It is clearly visible how the larger $\Delta x_{DEM}$ the larger the orange region (negative velocity) across the classifier rim, such unphysically enlarged region accelerates too strongly the small particles that eventually leave the milling chamber. Finally Figure B.1 show the discretization consequences on the tangential component of the fluid velocity profile and thus on the probability distribution of particle-particle and particle-wall collision velocities. Panel (a) shows the tangential velocity profile, the discretization does not affect significantly its behavior close to the classifier rim, however close to the vertical outer walls of the chamber the $\Delta x_{DEM} = 2.5\ mm$ profile experiences a sudden drop. Part of the mesh elements lay out of the milling chamber, where the fluid velocity is by definition set to zero, thus the average fluid velocity value associated to them drops significantly.

This phenomenon occurs recursively depending on the ratio between $\Delta x_{DEM}$ and the chamber diameter $D$, and the odd or even number of grid elements along the *xy* plane. It must be carefully avoided as the unphysical modification of the velocity profiles changes significantly the particle collision energy as highlighted in panel (b) of Figure B.2. For the other values of $\Delta x_{DEM}$ the velocity profile remains identical in shape with a slight shift to lower particle-wall collision velocities as $\Delta x_{DEM}$ increases. From this analysis $\Delta x_{DEM} = 1\ mm$ has been set for the DEM simulations presented in the whole paper.

A last point to be touched concerning DEM simulations is the accuracy of the triangular mesh representing the inner milling chamber walls. The region close to the walls is where most of the size reduction occurs by particle-particle and particle-wall collisions, it is thus mandatory to verify if a finer or a coarser mesh can influence the collision statistics. Figure B.3 (a) shows 4 different mesh portions named very fine (62.7K cells and nodes 31.4K nodes), fine (10.5K cells and nodes 5.2K nodes), moderate (10.0K cells and nodes 5.0K nodes) and coarse (6.9K cells and nodes 3.4K nodes). Each point represents a single particle-wall collision, color coded according to the collision velocity, coming from a multiple particle injection. The overall collision spatial distribution is very similar for all four meshes with a lower and an upper halo, however for the coarse end moderate meshes the halo is not uniform with the particles impacting predominantly on one corner of the large triangles. Simplifying a circular smooth surface with few large triangle results in a non-perfectly circular mesh, the impacted corners of the triangles are the ones protruding the most inside the milling chamber. For the fine and very fine meshes the density of collision points along the two halos is more uniform. Despite the irregular distribution of collision points introduced by the too coarse meshes the probability distribution for the collision velocity remains the same for all 4 cases, as visible in panel (b) of Figure B.3. For the DEM simulations presented in the paper we have selected a fine mesh.

Figures and captions

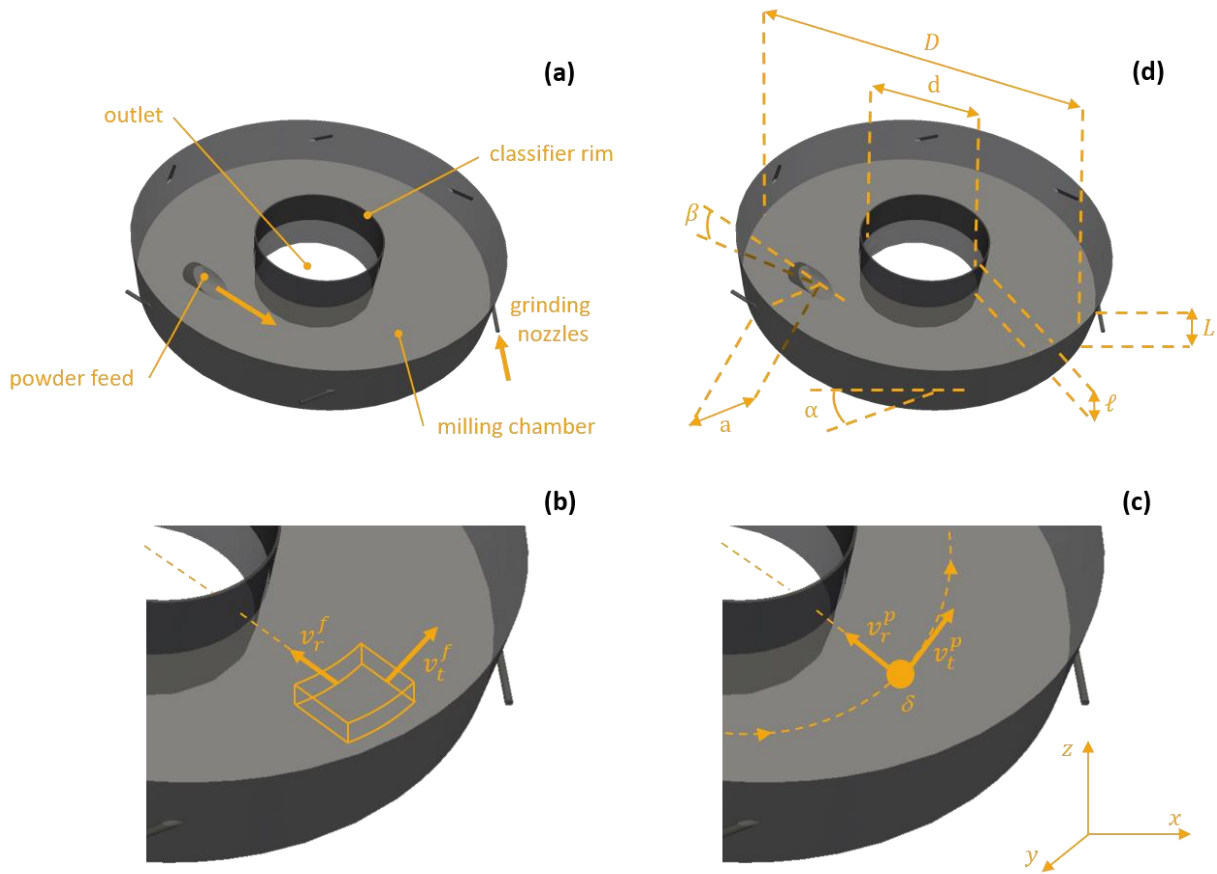

**Figure 1:** (a) Sketch of the model milling chamber geometry highlighting inlets, outlet and classifier. (b) velocity components in the Eulerian description of the milling fluid motion. (c) velocity components and trajectory in the Lagrangian description of the particle motion. (d) Principal geometric parameters characterizing the milling chamber.

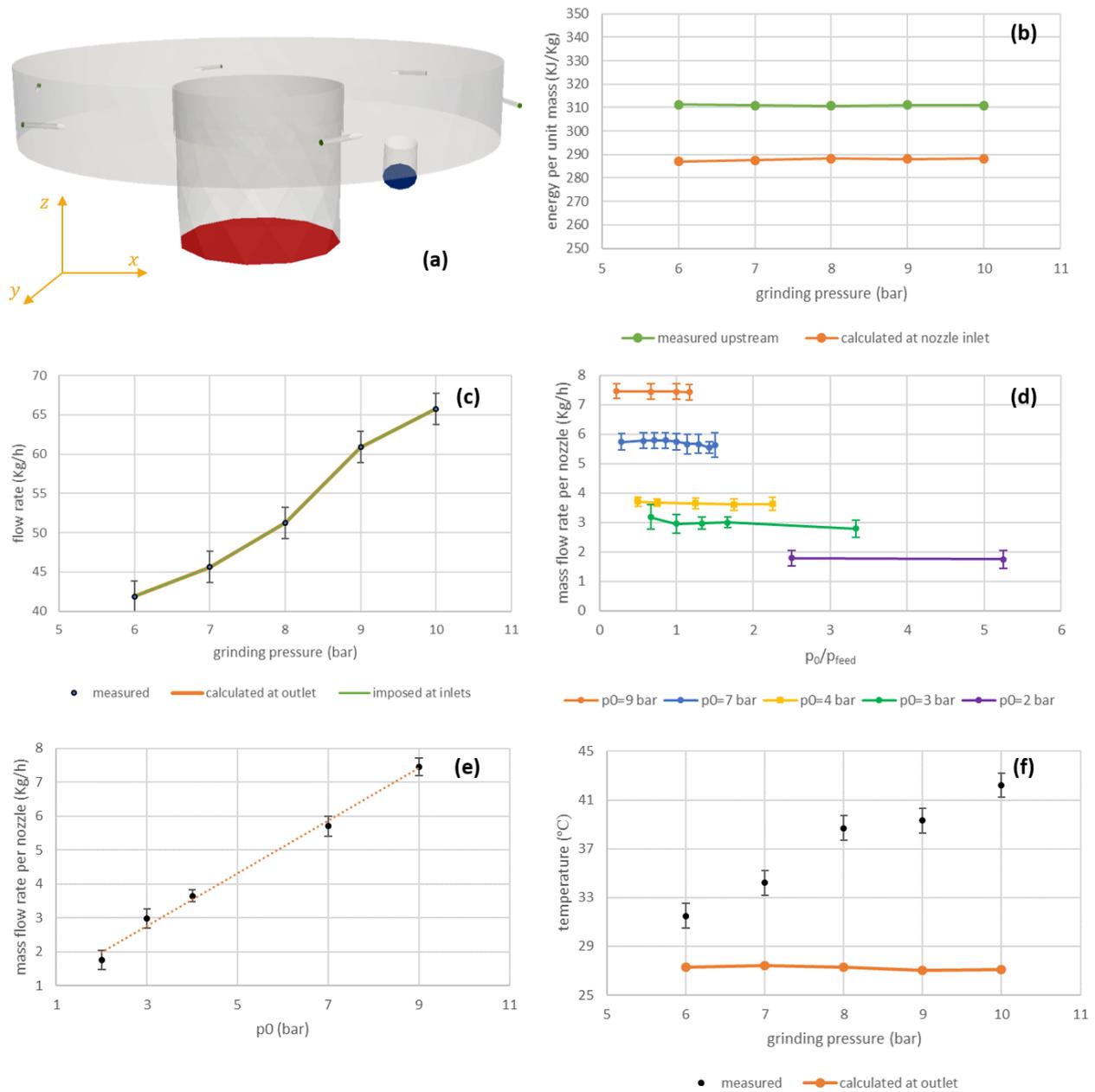

**Figure 2:** (a) sketch of the milling chamber highlighting inlets and outlet. (b) comparison between energy per unit mass measured upstream the nozzles and calculated at the nozzle entrance as a function of grinding pressure, the feed pressure is always one bar larger that grinding one, the outlet pressure is fixed at 1 atm. (c) milling fluid mass flow rate as a function of grinding pressure (feed and outlet pressure like in panel (b)): comparison between the one measured experimentally upstream the milling chamber, the one imposed at the inlets and the one found at the calculated at the outlet. (d) measured milling fluid mass flow rate as a function of the grinding to feed pressure ratio. (e) measured milling fluid mass flow rate as a function of the grinding pressure. (f) temperature of the outflowing milling gas as a function of grinding pressure (feed and outlet pressure like in panel (b)): comparison between the experimental measurements and calculated values.

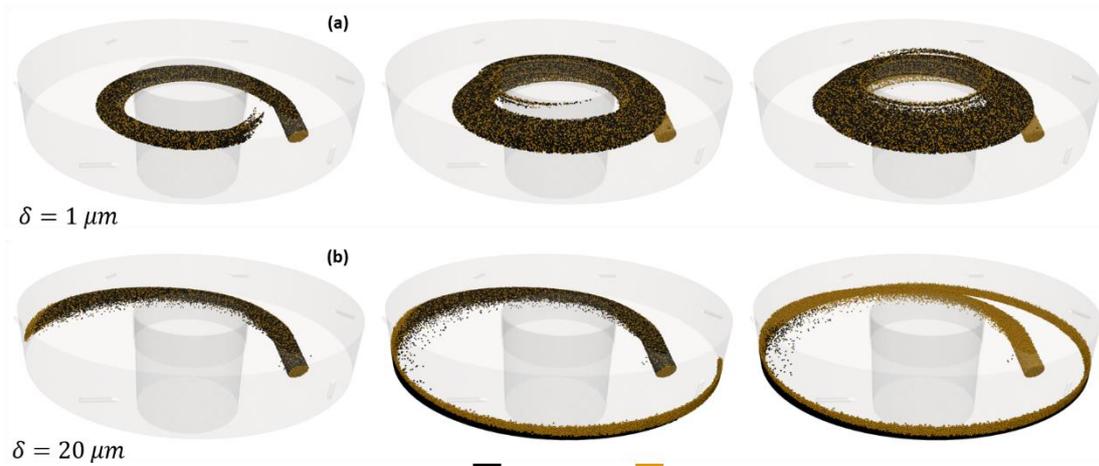

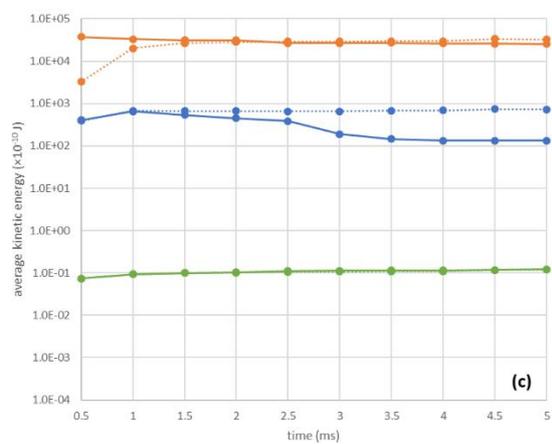
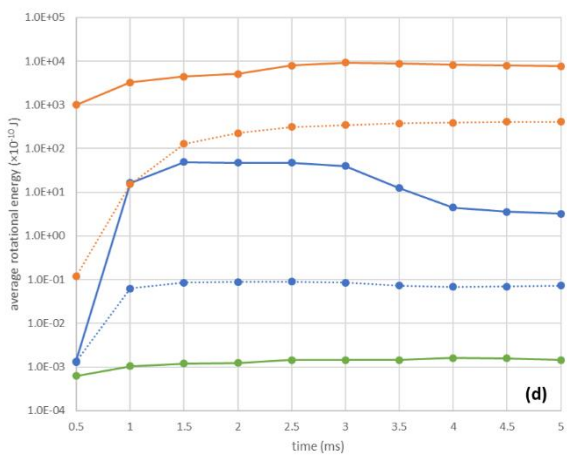

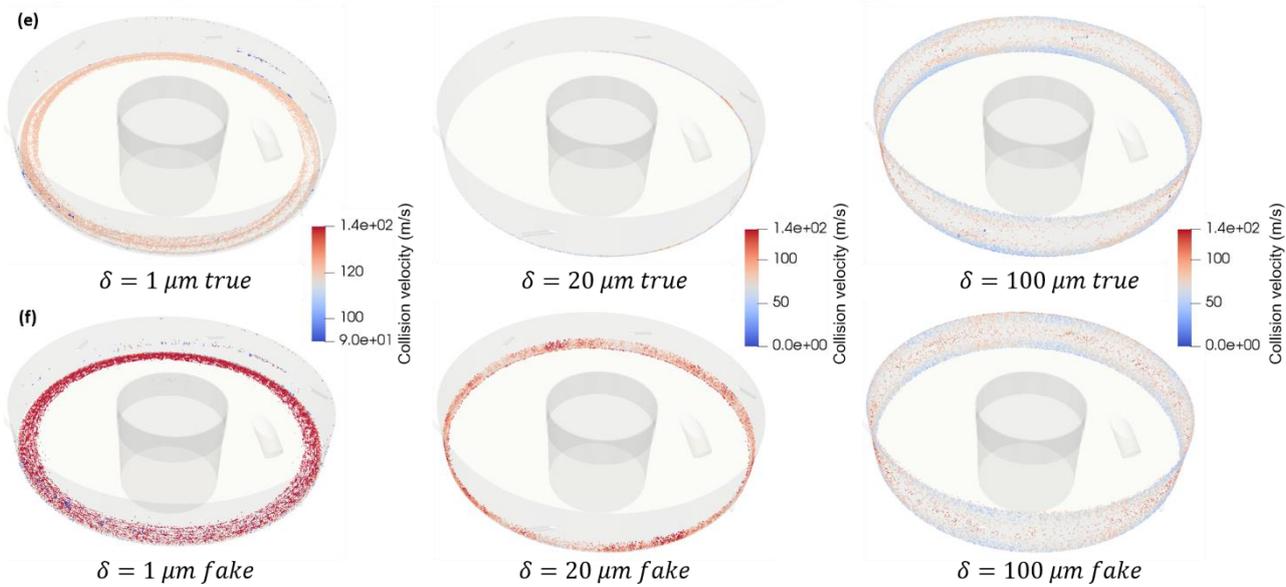

**Figure 3:** (a) and (b) subsequent snapshots superimposing the DEM simulations of true (black) and fake (orange) density/diameter cases for $\delta = 1$ and $20\ \mu m$ respectively. The CFD steady state is obtained with $p_0 = 7\ bar,\ p_{feed} = 8\ bar,\ p_{out} = 1\ atm, \alpha = 26°, d = 35\ mm,\ \ell = 9.5\ mm$. (c) and (d) per particle kinetic and rotational energy as a function of time for three different DEM simulation in which 1, 20 and $100\ \mu m$ particles have been injected both using the true and fake diameter methods. The curve for the rotational energy of the case $\delta = 1\ \mu m$ with fake diameter has been omitted as the values where comparable to the numerical error, it can thus be considered zero. (e) and (f) simulation snapshots displaying the location of particle-wall collisions for $\delta = 1, 20$ and $100\ \mu m$ in both the true and fake cases, the color code represents the collision velocity and is the same for every true/fake couple in order to help visualizing the differences between the two methods.

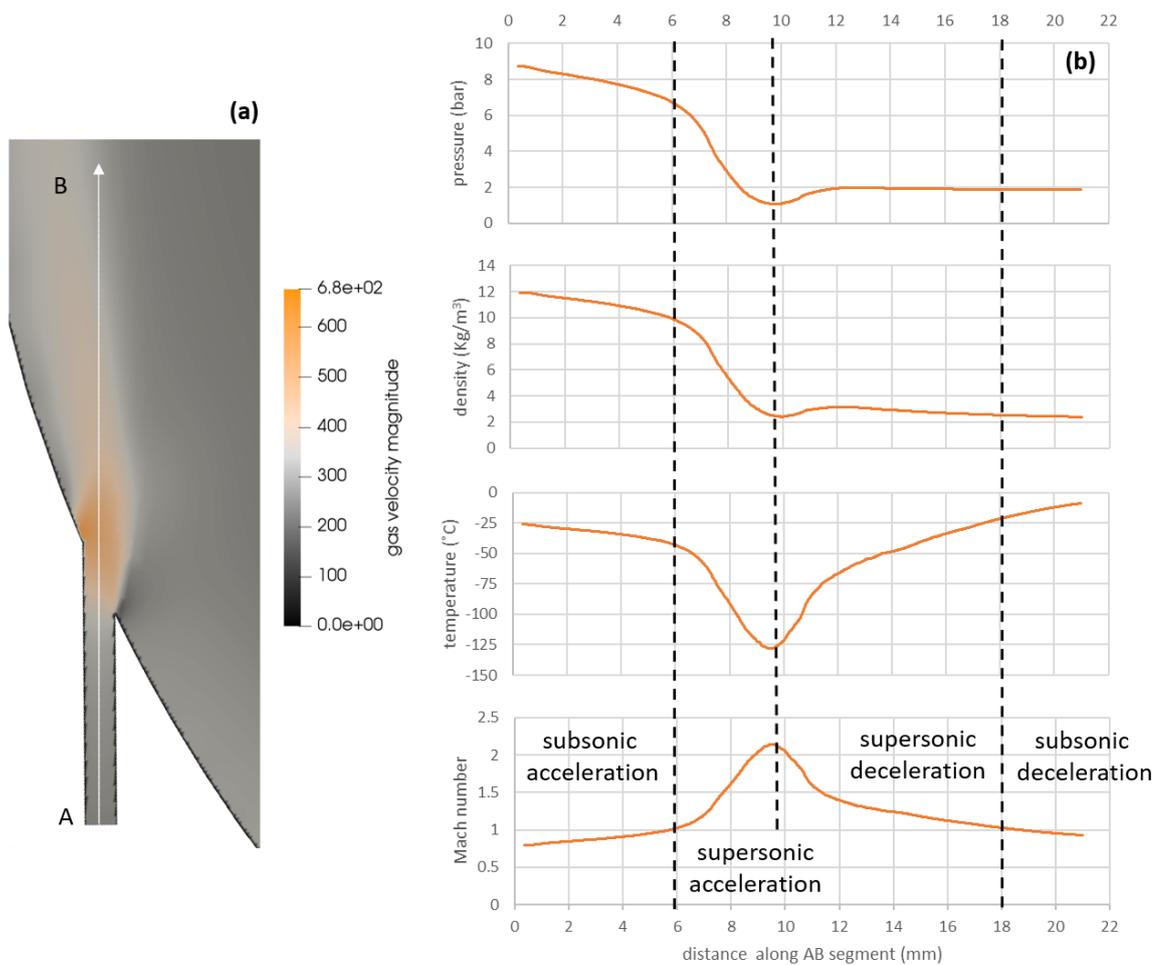

**Figure 4:** Fluid behavior at nozzles for the case $p_0 = 10 \, bar$, $p_{feed} = 11 \, bar$. (a) map of the milling fluid velocity magnitude inside and in the proximity of a grinding nozzle. (b) Pressure, density, temperature and Mach number values along the AB segment of panel (a) moving from point A to B. The dashed lines separate four regions.

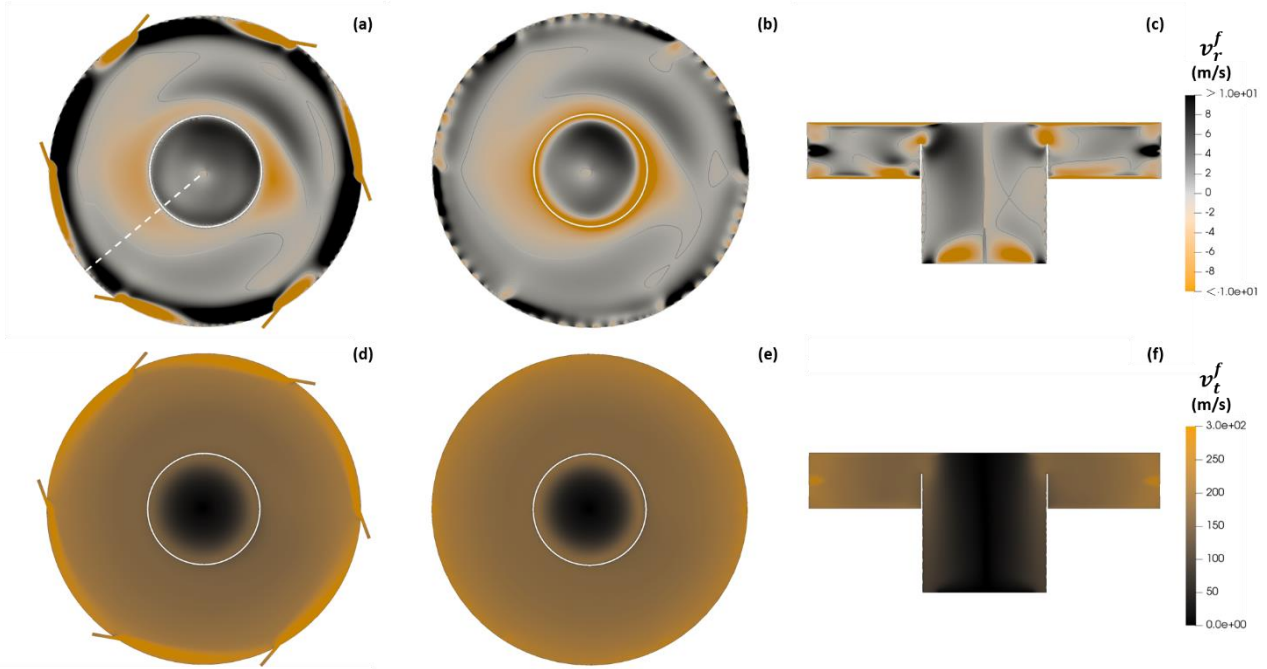

**Figure 5:** Fluid velocity components in the grinding chamber for the case $p_0 = 8 \, bar$, $p_{feed} = 9 \, bar$, $\alpha = 26°$, $d = 35 \, mm$ and $\ell = 9.5 \, mm$. (a), (b) and (c) represent the radial component of the fluid velocity at the nozzle plane, at the classifier rim plane and along a plane perpendicular to the chamber diameter respectively. The plotting planes are displayed in Figure 6 (a) in orange, blue and green respectively. The thin lines in these panels are the $v_r^f = 0$ isolines, they highlight the regions where the velocity approach zero and change sign. The white thick dashed line represents the direction along which the plots of Figure 6 have been taken, i.e. the line on which the AB and CD vectors of Figure 6 (a) lay. Panels (d), (e) and (f) show the tangential component of the fluid velocity in the same plotting planes of the other panels.

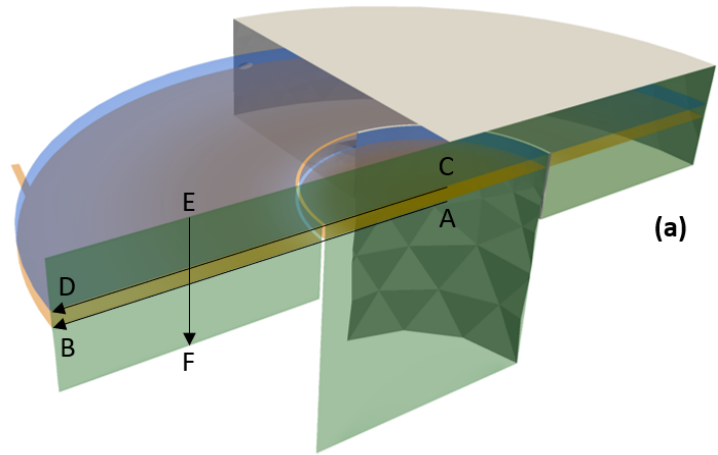

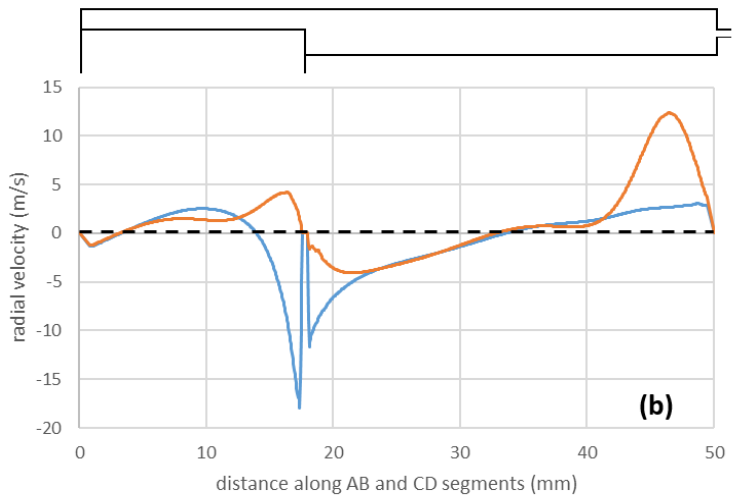
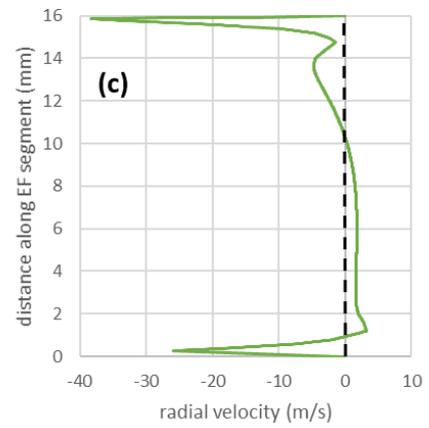

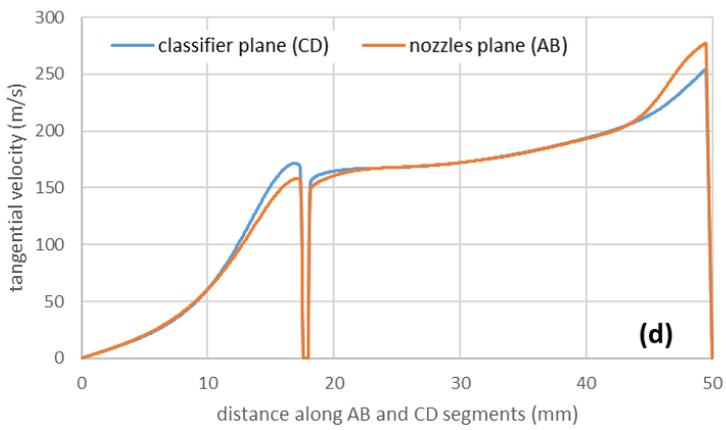
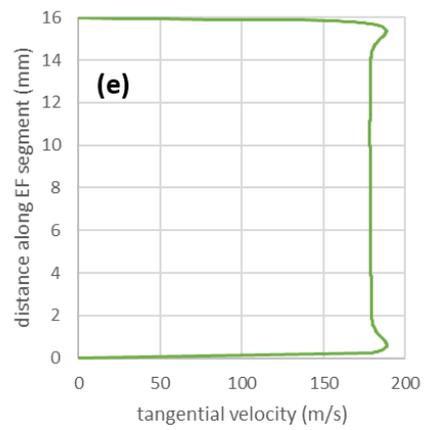

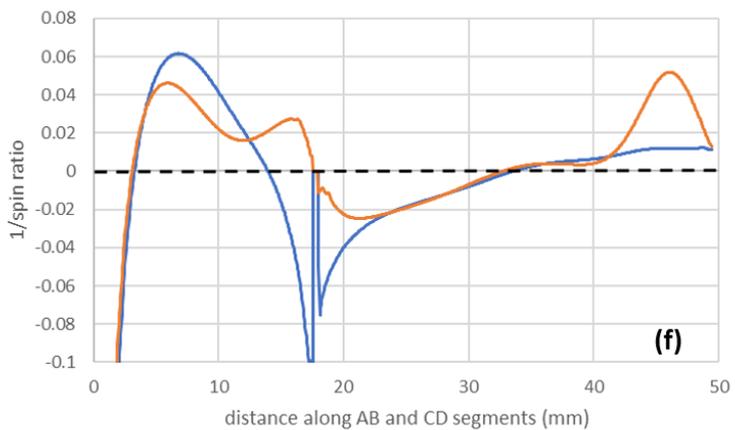
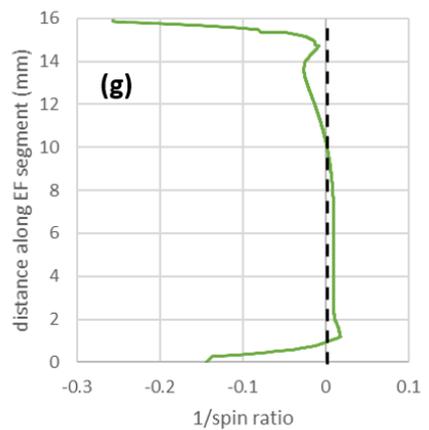

**Figure 6:** 2D plots of fluid velocity components for the case $case\ p_0 = 8\ bar,\ p_{feed} = 9\ bar,\ \alpha = 26°, d = 35\ mm\ and\ \ell = 9.5\ mm$. (a) milling chamber cross section showing the plotting planes and lines of figures 4 and 5. (b) to (g) plots of the radial and tangential components of the fluid velocity and plots of their ratio, the spin ratio, along the three directions AB, CD and EF described in panel (a).

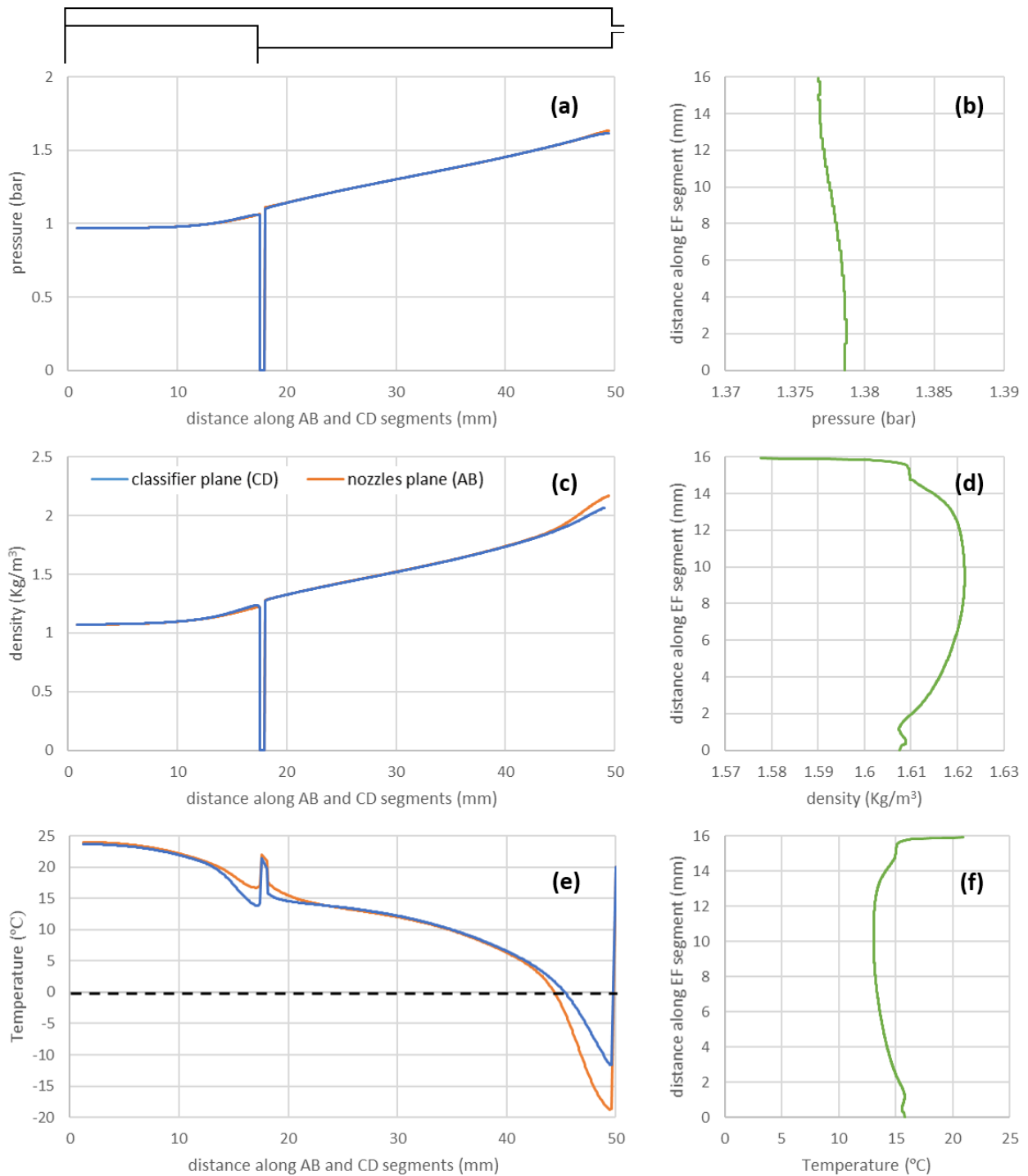

**Figure 7:** 2D plot of the thermodynamic scalar variables of the fluid along the same lines of Figure 6 (a). (a) and (b) fluid pressure, (c) and (d) fluid density, (e) and (f) fluid temperature.

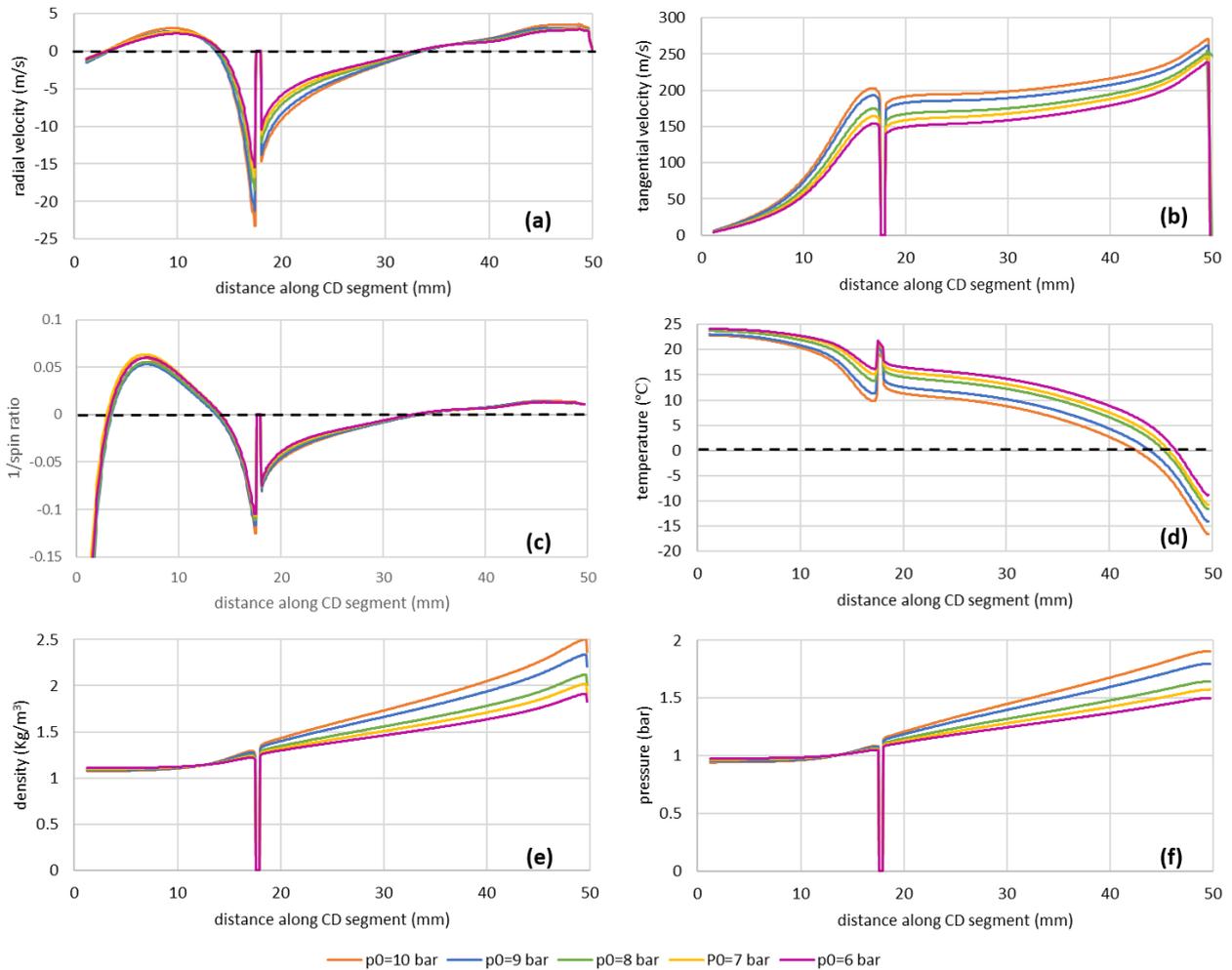

**Figure 8:** 2D plots of fluid velocity components and thermodynamic variables for different grinding pressures in the case $\alpha = 26°, d = 35\ mm, \ell = 9.5\ mm$ and $p_{out} = 1\ atm$. (a) and (b) radial and tangential components of the fluid velocity, (c) inverse of the spin ratio, (d), (e) and (f) temperature, density and pressure respectively calculated along the CD segment.

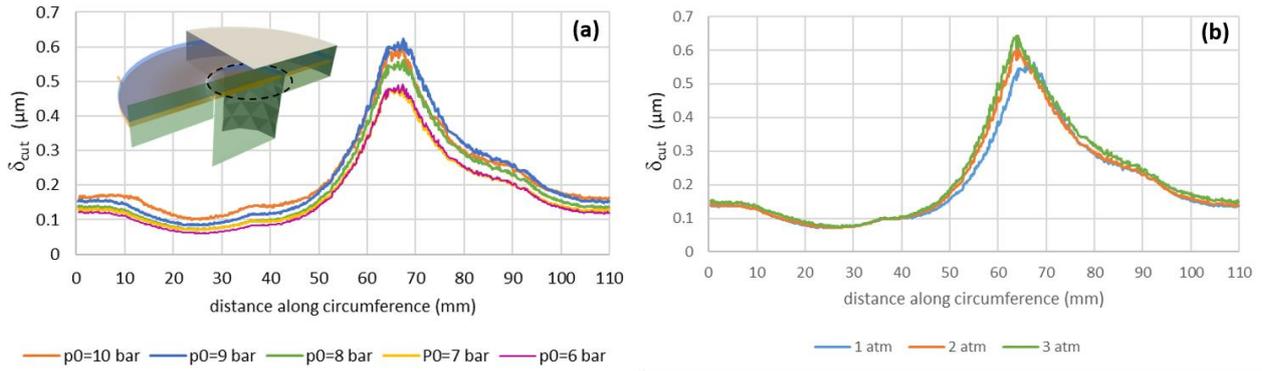

**Figure 9:** (a) cut size along the classifier circumference for different grinding pressures in the case $\alpha = 26°, d = 35\ mm, \ell = 9.5\ mm$ and $p_{out} = 1\ atm$. (b) cut size along the classifier circumference for different outlet pressures, feeding and geometry parameters like in panel (a). The inset of panel (a) represent the circumference above the classifier rim on which the calculation has been done.

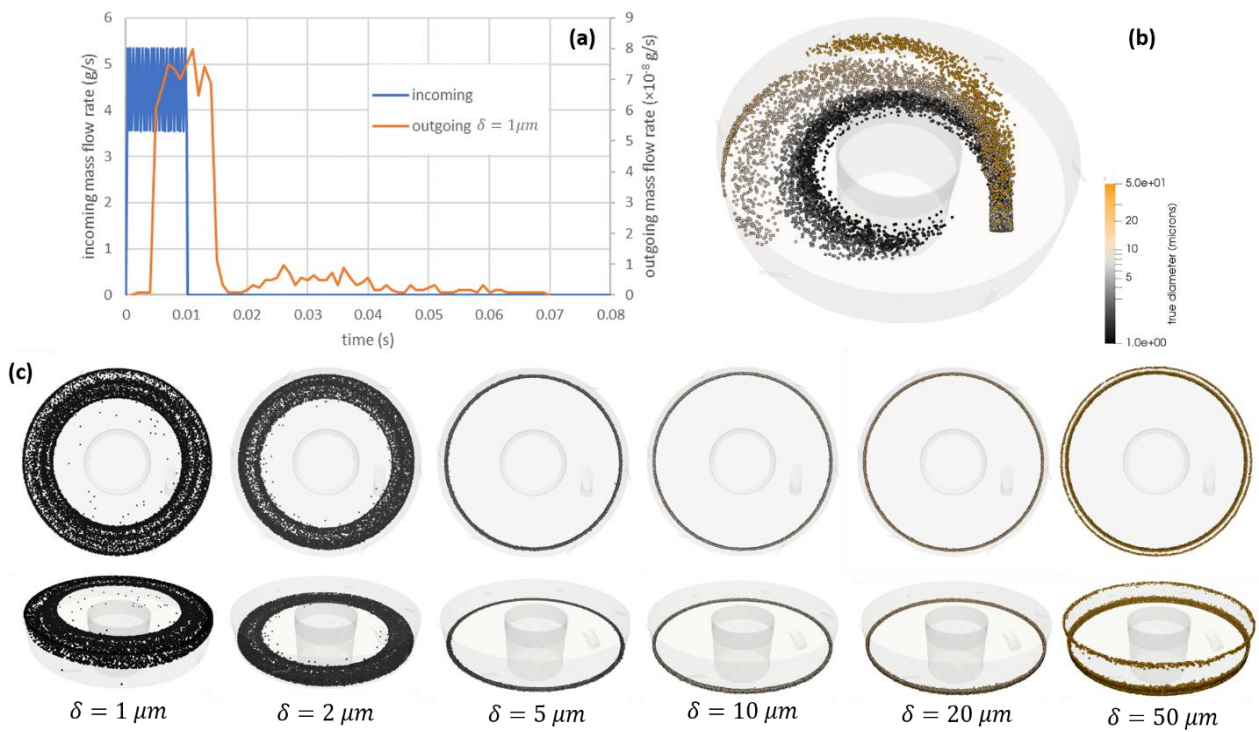

**Figure 10:** Poly-disperse particle injection for the case $p_0 = 7\ bar, \alpha = 26°, d = 35\ mm, \ell = 9.5\ mm$ and $p_{out} = 1\ atm$. (a) incoming mass flow rate from the feed inlet (blue) and outgoing mass flow rate from the classifier (orange) during the DEM simulation. (b) particle positions few moments after the injection started. (c) top and side view of particle positions at 0.08 s, i.e. once a steady state is fully developed, all the particle

diameters are simultaneously present in the milling chamber but they have been plotted separately for convenience. The color code is the same for both panels (b) and (c).

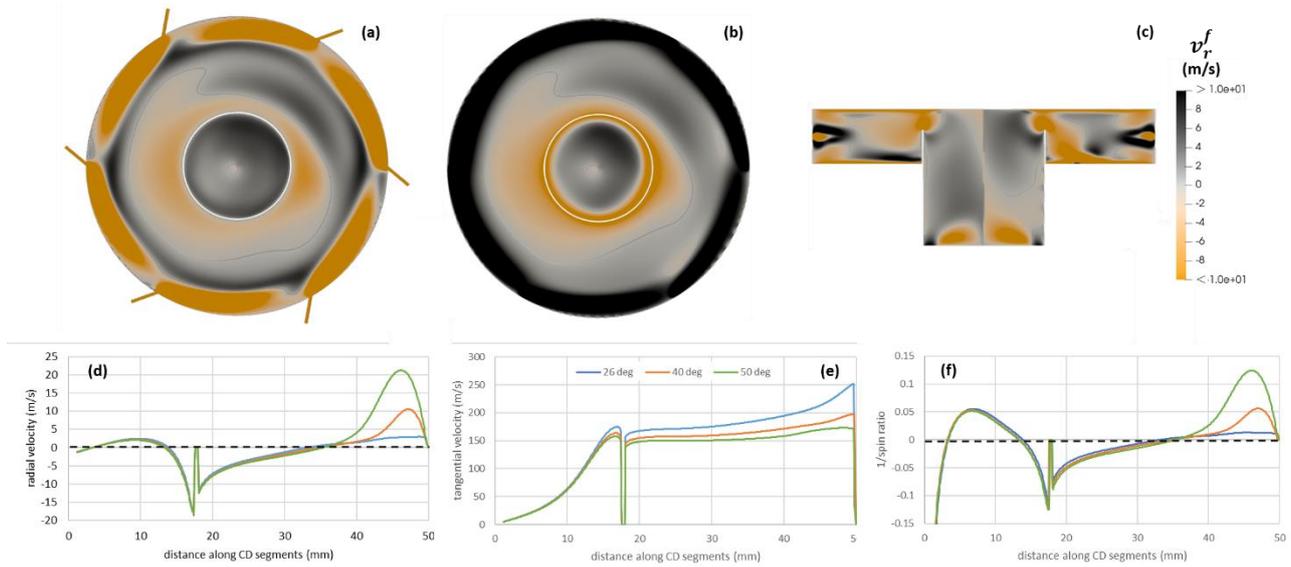

**Figure 11:** Milling fluid behaviour as a function of the nozzle angle $\alpha$ for the case $p_0 = 8\ bar$, $p_{feed} = 9\ bar$, $d = 35\ mm$ and $\ell = 9.5\ mm$. (a) to (c) radial component of fluid velocity, at the nozzle plane, at the classifier rim plane and along a plane perpendicular to the chamber diameter respectively. The plotting planes are displayed in Figure 6 (a) in orange, blue and green respectively. The thin lines in these panels are the $v_r^f = 0$ isolines. (d) and (e) show the radial and tangential components of the fluid velocity along the CD segment moving from the center to the periphery of the milling chamber. (f) show the spin ratio along the same direction.

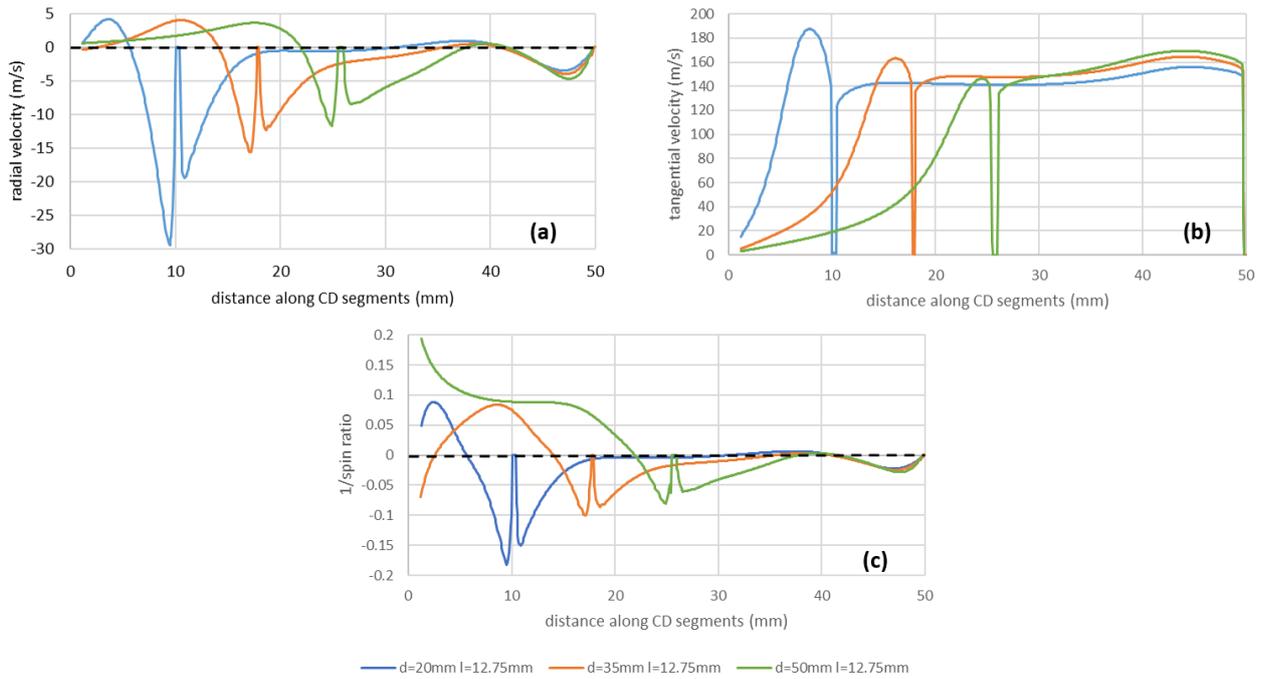

**Figure 12:** Velocity profiles as function of the classifier diameter $d$ for the case $p_0 = 8\ bar$, $p_{feed} = 9\ bar$, $\alpha = 50°$, $\ell = 12.75\ mm$. (a) and (b) show the radial and tangential components of the fluid velocity along the CD segment moving from the center to the periphery of the milling chamber for the case . (f) show the spin ratio along the same direction.

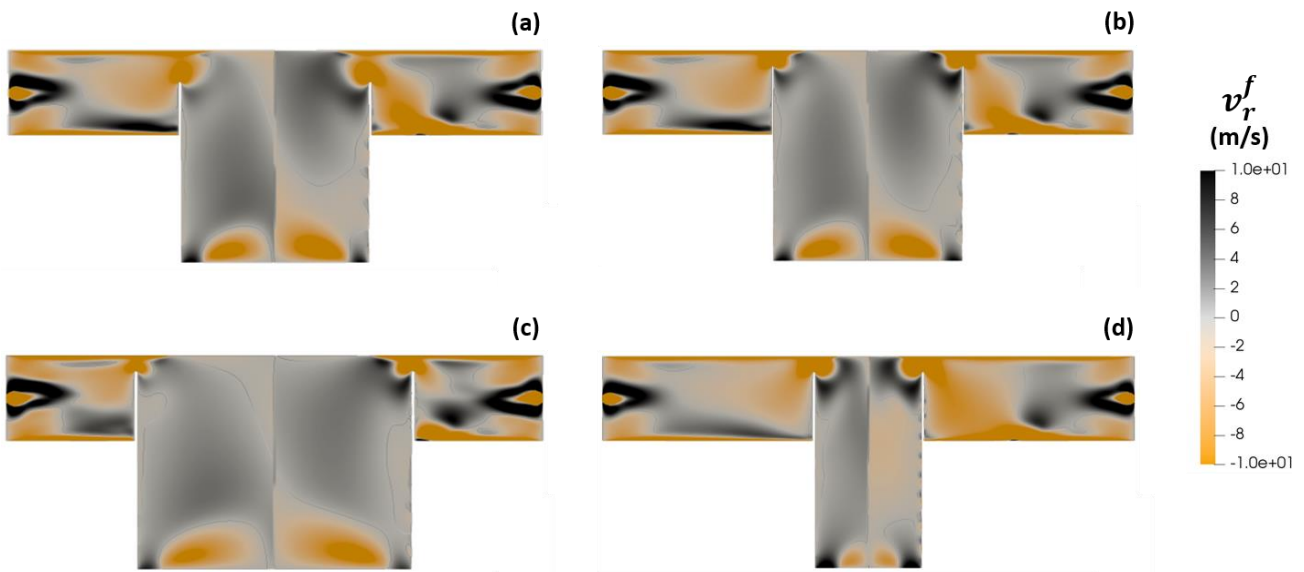

**Figure 13:** Radial velocity maps in a plane perpendicular to the milling chamber disk, i.e. on the green plane of Figure 6 (a), for the case $p_0 = 8\ bar$, $p_{feed} = 9\ bar$, $\alpha = 50°$ and (a) $d = 35\ mm$, $\ell = 9.5\ mm$; (b) $d =$

$35\ mm$, $\ell = 12.75\ mm$; (c) $d = 50\ mm$, $\ell = 12.75\ mm$; (d) $d = 20\ mm$, $\ell = 12.75\ mm$. The thin lines in the panels are the $v_r^f = 0$ isolines.

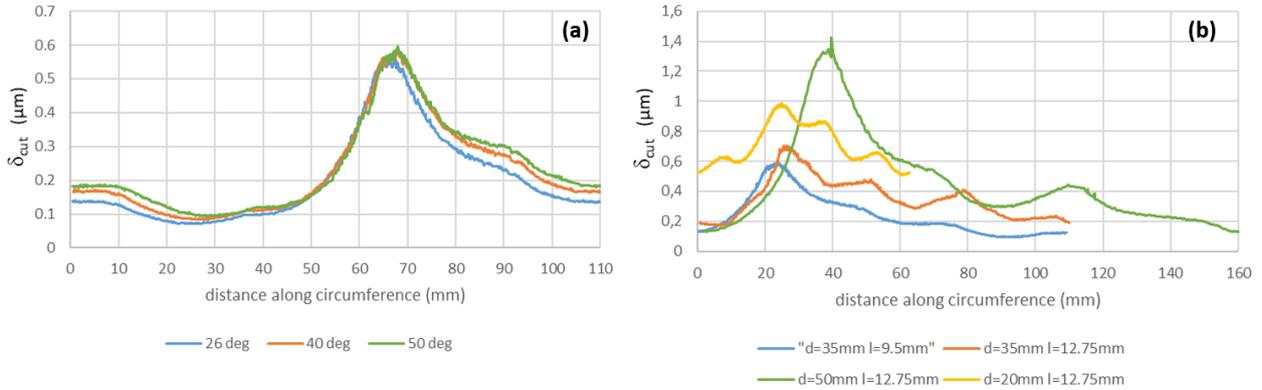

**Figure 14:** (a) cut size along the classifier circumference for different nozzle angles $\alpha$ for the case $p_0 = 8\ bar$, $p_{feed} = 9\ bar$, $d = 35\ mm$, $\ell = 9.5\ mm$ and $p_{out} = 1\ atm$. (b) cut size along the classifier circumference for different classifier geometries for the case $p_0 = 8\ bar$, $p_{feed} = 9\ bar$ and $p_{out} = 1\ at$.

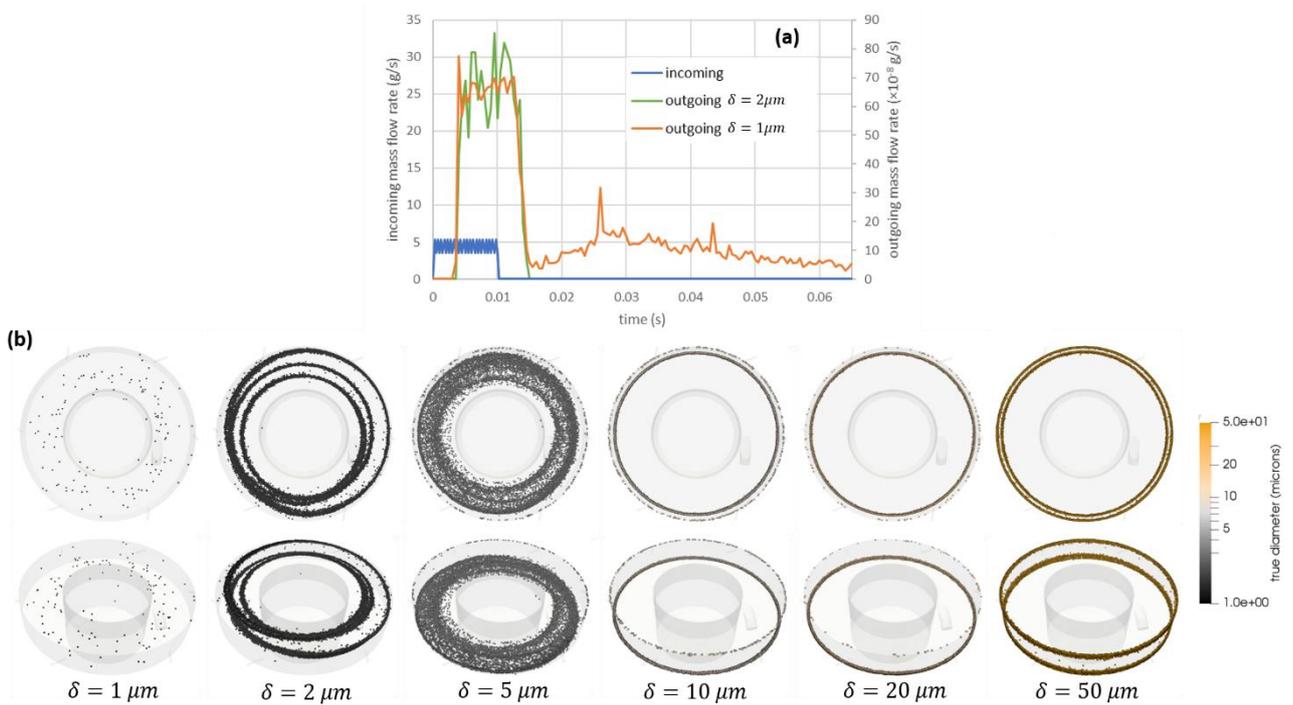

**Figure 15:** Poly-disperse particle injection for the case $p_0 = 8\ bar$, $p_{feed} = 9\ bar$, $p_{out} = 1\ atm$ and with geometric parameters $\alpha = 50°$, $d = 50\ mm$, $\ell = 12.75\ mm$. (a) incoming mass flow rate from the feed inlet (blue) and outgoing mass flow rate from the classifier (orange and green) during the DEM simulation. (b) top

and side view of particle positions at 0.08 s, i.e. once a steady state is fully developed, all the particle diameters are simultaneously present in the milling chamber but they have been plotted separately for convenience.

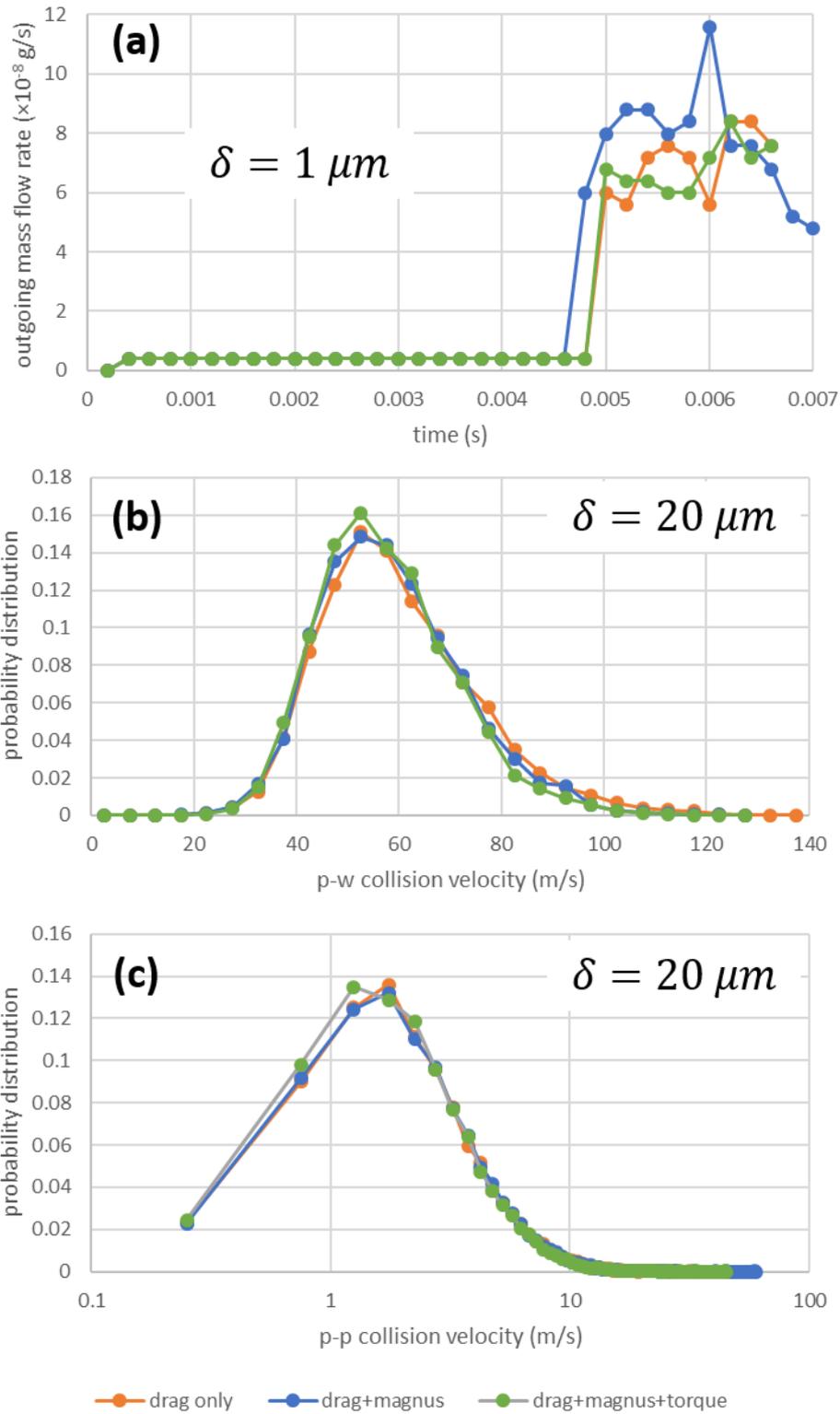

**Figure 16:** Mono-disperse particle injection for the case $p_0 = 8\ bar$, $p_{feed} = 9\ bar$, $p_{out} = 1\ atm$ and with geometric parameters $\alpha = 50°, d = 50\ mm, \ell = 12.75\ mm$. (a) outgoing mass flow rate from the classifier of 1 μm particles when only the drag force is applied (orange), drag and Magnus lift are applied (blue) and drag, Magnus and Stokes torque are applied (green). (b) and (c) probability distributions for particle-wall and particle-particle collision velocity for 20 μm diameter particles in the same three possible choices of particle-fluid interaction of panel (a).

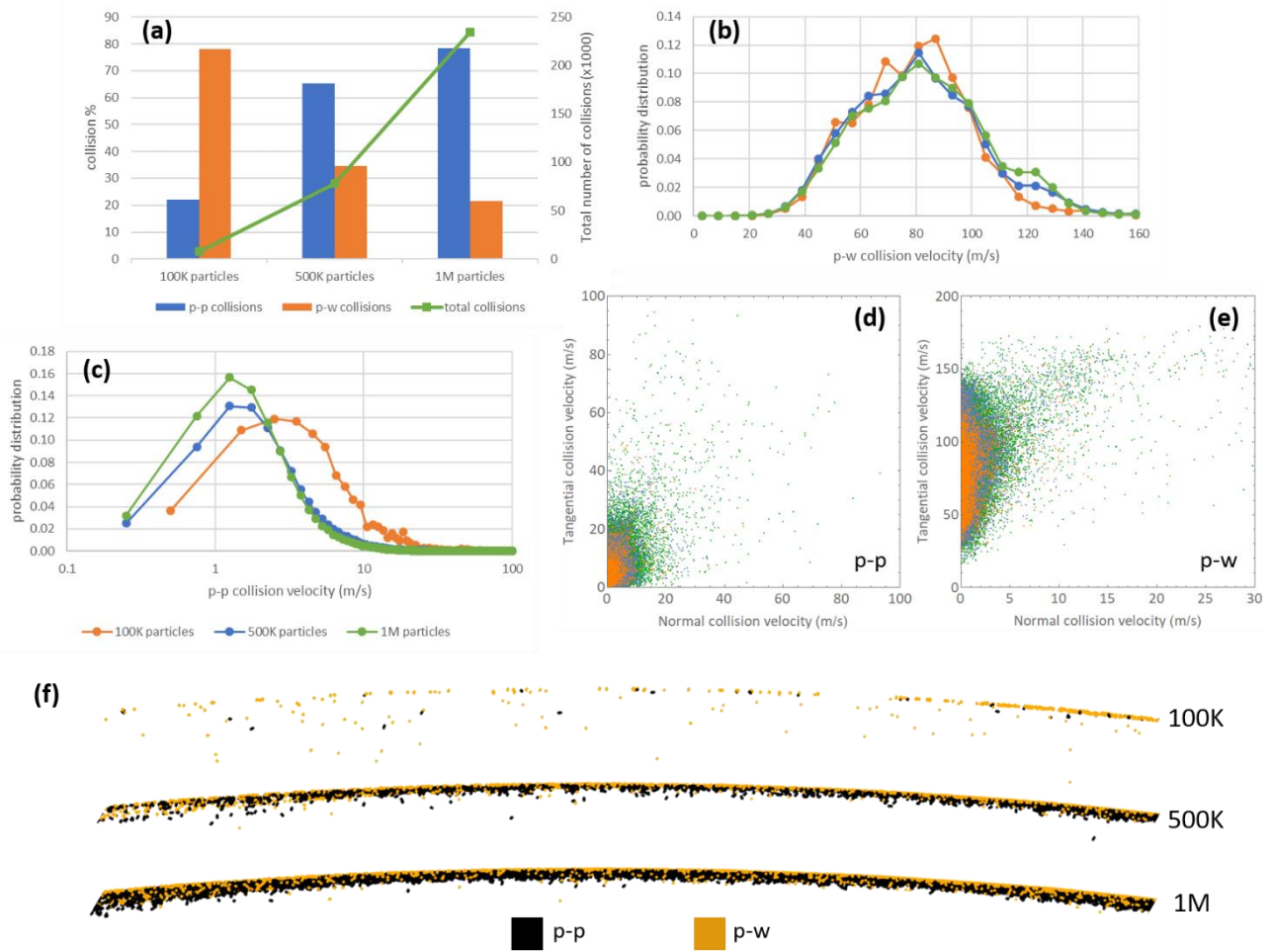

**Figure 17:** Mono-disperse 20 μm diameter particle injections for the case $p_0 = 8\ bar$, $p_{feed} = 9\ bar$, $p_{out} = 1\ atm$ and with geometric parameters $\alpha = 50°, d = 50\ mm, \ell = 12.75\ mm$. (a) total number of collisions as a function of the number of injected particles (green line with right vertical axis) and percentage of particle-wall and particle-particle collisions for every injection (histogram with left vertical axis). (b) and (c) probability distributions for particle-wall and particle-particle collision velocity for 100k, 500K and 1M particle injections. (d) and (e) scatter plots representing, for each particle-particle or particle-wall collision, the normal and tangential components of the collision velocity. (f) maps of the particle-particle and particle-wall collisions as a function of the number of injected particles.

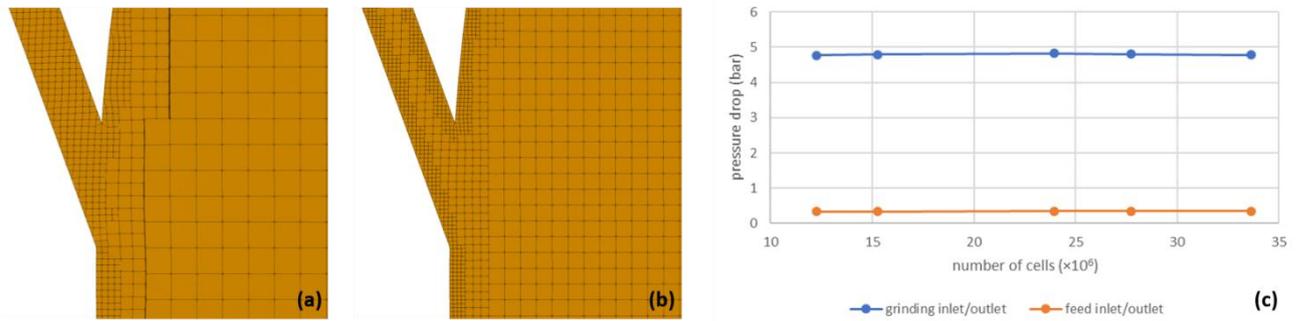

**Figure A.1:** Mesh sensitivity analysis for the case $p_0 = 7\ bar$, $p_{feed} = 8\ bar$, $\alpha = 26°, d = 35\ mm$ and $\ell = 9.5\ mm$. Panels (a) and (b) show a detail of the CFD mesh close to a grinding nozzle using a 12 and 34 million elements mesh. (c) pressure drops calculated between grinding nozzle and outlet and between feed inlet and outlet as a function of the mesh size

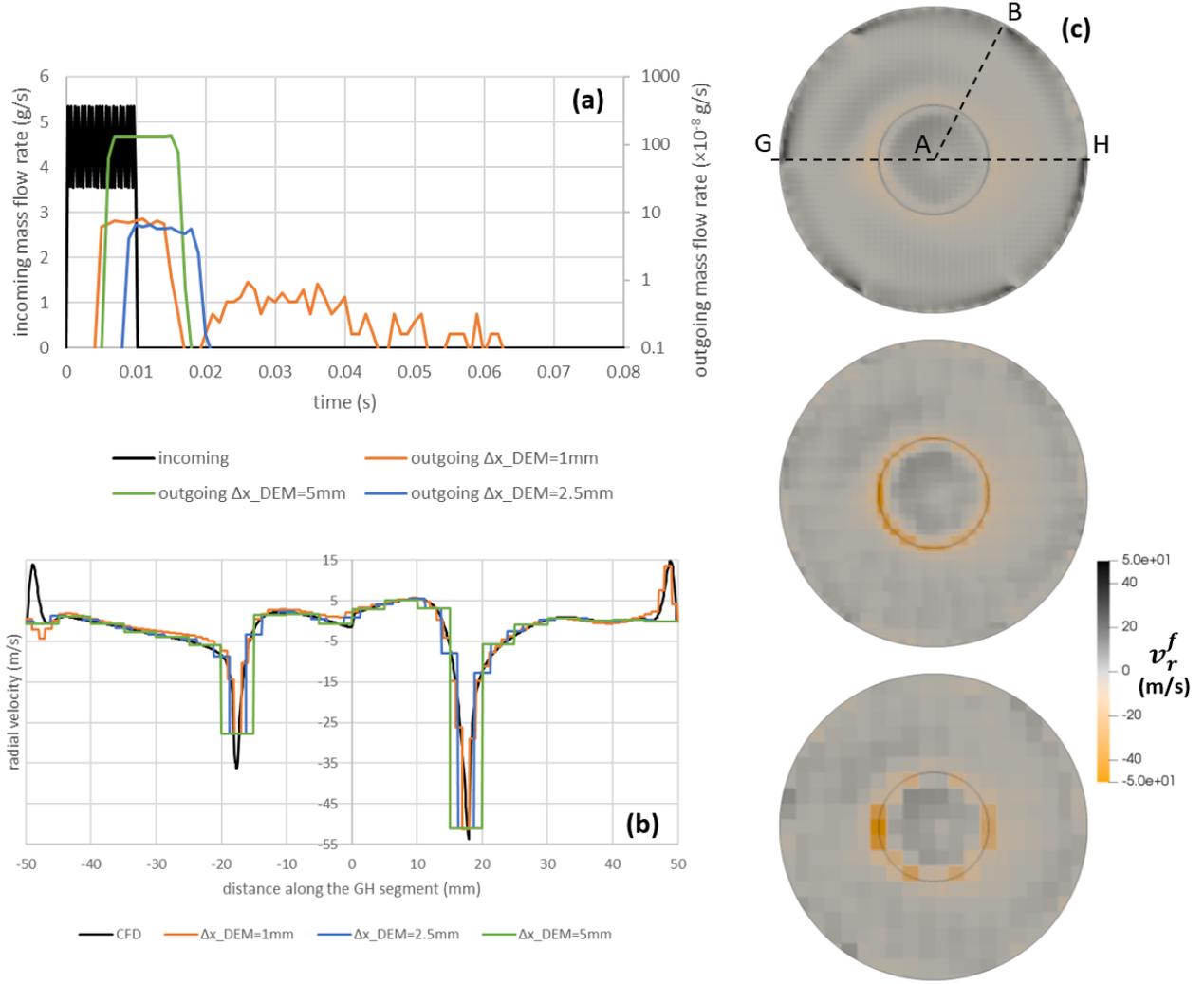

**Figure B.1:** effect of the DEM mesh size on particle classification for the case $p_0 = 7\ bar$, $p_{feed} = 8\ bar$, $p_{out} = 1\ atm$, $\alpha = 26°$, $d = 35\ mm$, $\ell = 9.5\ mm$. (a) incoming mass flow rate from the feed inlet (black) and outgoing mass flow rate from the classifier (coloured curve) during the DEM simulation with different $\Delta x_{DEM}$ values, all the classified particles have $\delta = 1\mu m$. (b) radial velocity profile along the GH line shown in panel (c) from the CFD calculation (black) and after resampling on coarser meshes with different $\Delta x_{DEM}$ values. (c) radial velocity maps on the classifier rim plane for the three different $\Delta x_{DEM}$ values. The AB line, on which the CFD velocity profiles are plotted in the rest of the paper, is also shown to highlight the different orientation with respect to GH.

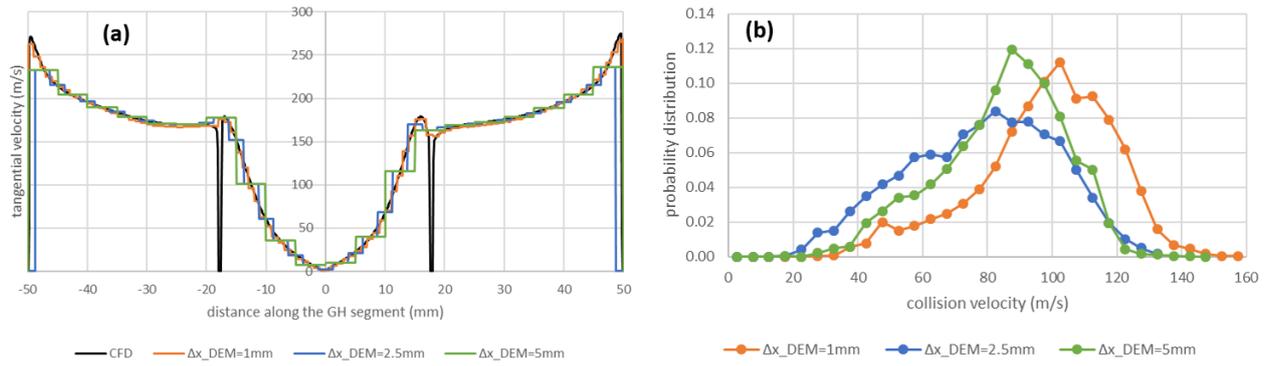

**Figure B.2:** effect of the DEM mesh size on particle-wall collisions for the same case of Figure B.1. (a) tangential velocity profile along the GH line, shown in panel (c) of Figure B.1, from the CFD calculation (black) and after resampling on coarser meshes with different $\Delta x_{DEM}$ values. (b) probability distribution of collision velocities for 100 $\mu m$ particles against vertical grinding chamber walls for different $\Delta x_{DEM}$ values.

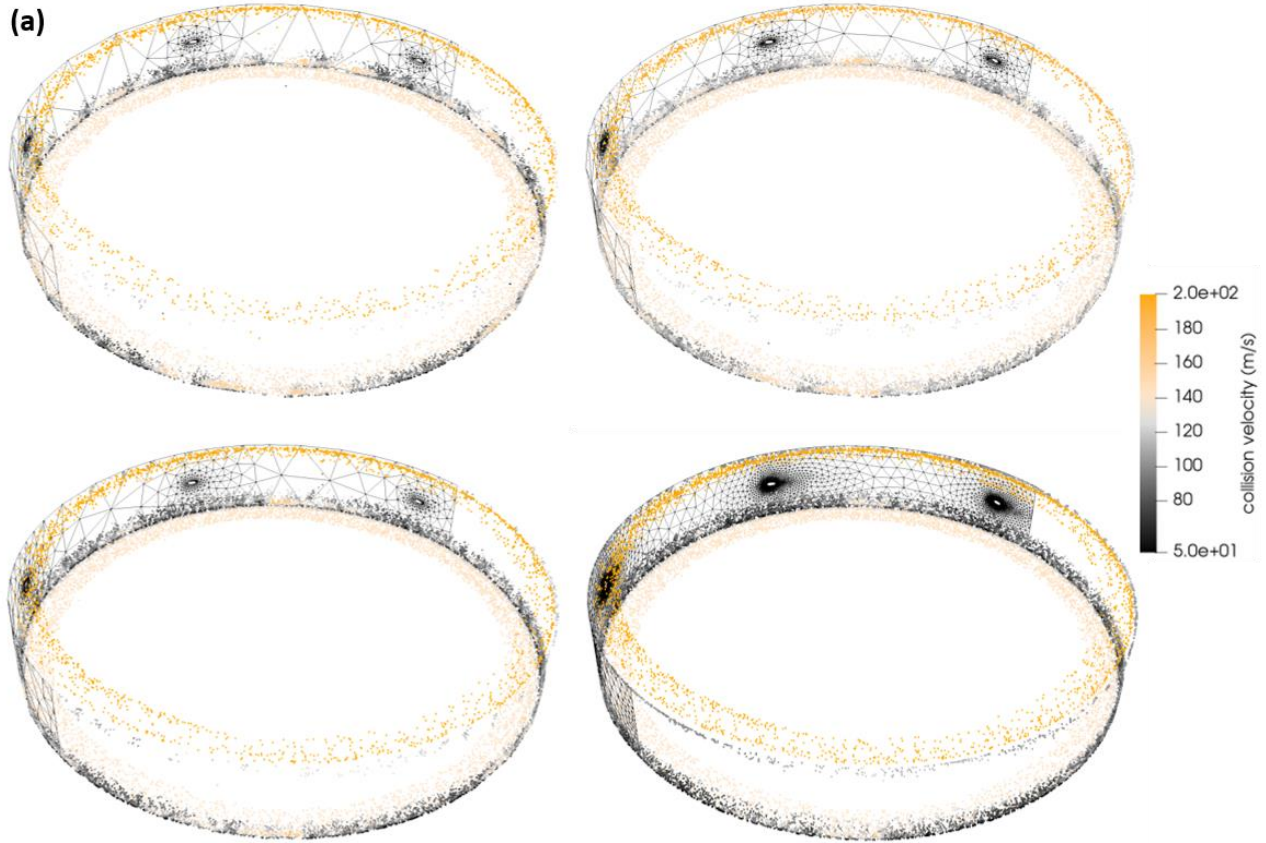

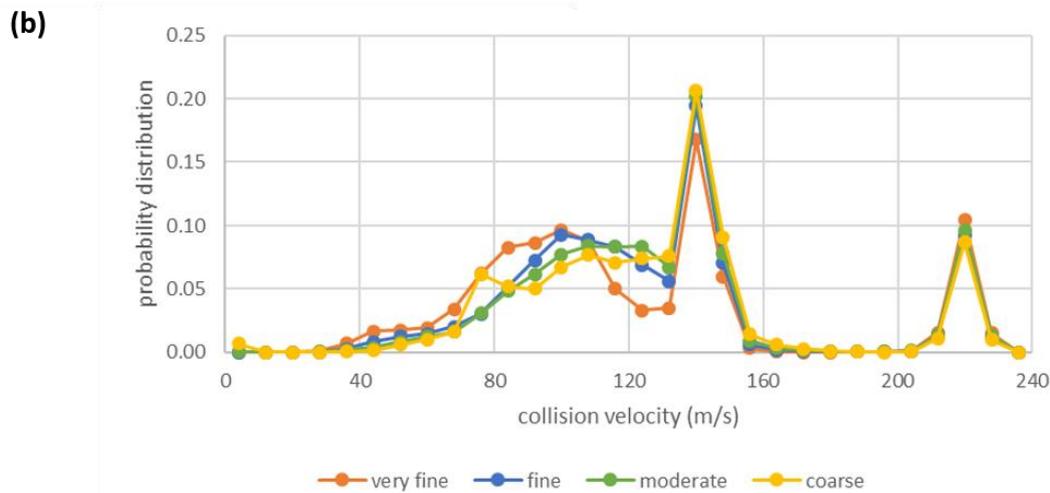

**Figure B.3:** Effect of the DEM triangular mesh accuracy for the same case of Figure B.1. (a) portion of the DEM mesh for coarse, moderate, fine and very fine meshes respectively from top left to bottom right. The dots represent single particle-wall collision events occurring in a single time instant coloured according to the impact velocity. (b) probability distribution of collision velocity averaged over many time instants once the DEM steady state is reached for the four meshes of panel (a).

List of symbols

| Symbol | Meaning | Units/Dimensions |
|---|---|---|
| $p_0$ | Absolute/relative grinding pressure upstream the nozzles | bar or barg |
| $p_{feed}$ | Absolute/relative powder feed pressure | bar or barg |
| $T_0$ | Upstream milling fluid temperature | °K or °C |
| $T_t$ | Milling fluid temperature at the nozzle throat | °K or °C |
| $T_{out}$ | Milling fluid temperature at the chamber outlet | °K or °C |
| $\dot{m}_{max}$ | Grinding flow rate | Nm$^3$/h or Kg/h |
| $\dot{m}_{feed}$ | Feed flow rate | Nm$^3$/h or Kg/h |
| $A_t$ | Nozzle cross sectional area | m$^2$ |
| $M$ | Milling fluid molar mass | Kg/mol |
| $\gamma$ | Specific heat ratio | / |
| $R$ | Ideal gas constant | J/(°K mol) |
| $v_t$ | Milling fluid velocity at the nozzle throat | m/s |
| $v_t^p$ | Particle tangential velocity component | m/s |
| $v_r^p$ | Particle radial velocity component | m/s |
| $v_t^f$ | Fluid tangential velocity component | m/s |
| $v_r^f$ | Fluid radial velocity component | m/s |
| $\overrightarrow{v^p}$ | Particle velocity vector | m/s |
| $\overrightarrow{v^f}$ | Fluid velocity vector field | m/s |
| $\overrightarrow{r^p}$ | Particle position vector | m |
| $\overrightarrow{\omega^p}$ | Particle angular velocity | rad/s |
| $\overrightarrow{\omega^f}$ | Fluid angular velocity | rad/s |
| $Stk$ | Stokes number | / |
| $\rho_p$ | Particle density | Kg/m$^3$ |
| $\rho_f$ | Milling fluid density | Kg/m$^3$ |
| $\mu$ | Milling fluid dynamic viscosity | N·s/m$^2$ |
| $v_0$ | Milling fluid free stream velocity | m/s |
| $t_p$ | Particle residence time in the milling chamber | s |
| $m_h$ | Powder hold-up mass | Kg |
| $n$ | Powder volume fraction | / |
| $V_p$ | Chamber volume occupied by powder | m$^3$ |
| $V_f$ | Chamber volume occupied by the fluid | m$^3$ |
| $V$ | Generic milling fluid volume | m$^3$ |
| $T$ | Generic milling fluid temperature | °K or °C |
| $p$ | Generic milling fluid pressure | bar or barg |
| $Re$ | Reynolds number | / |
| $C_D$ | Drag coefficient | / |
| $C_L$ | Lift coefficient | / |
| $\delta$ | Particle diameter | m |
| $\delta_{cut}$ | Particle cut size | m |
| $r$ | Particle orbit radius | m |
| $E$ | Particle Young modulus | N/m$^2$ |
| $\nu$ | Poisson ration | / |
| $H$ | Particle hardness | N/m$^2$ |
| $K_C$ | Fracture toughness | N/m$^{3/2}$ |
| $u$ | Milling fluid energy per unit mass | J/Kg |

| Symbol | Description | Units |
|---|---|---|
| $e$ | Milling fluid internal energy density | J/m³ |
| $h$ | Milling fluid enthalpy per unit mass | J/Kg |
| $Re$ | Particle Reynolds number | / |
| $v$ | Generic fluid velocity module | m/s |
| $p_{out}$ | Absolute outlet pressure | bar or atm |
| $\epsilon$ | Rate of dissipation of turbulent kinetic energy | J/(Kg s) |
| $k$ | Turbulent kinetic energy density | J/Kg |
| $q$ | Thickness of the fluid boundary layer in the proximity of walls | m |
| $\Delta x_{CFD}$ | Characteristic CFD mesh size | m |
| $\Delta x_{DEM}$ | Characteristic DEM mesh size | m |
| $\Delta t$ | Timestep for the integration of the particle equation of motion | s |